\documentclass{sig-alternate}

\usepackage{graphicx}
\usepackage{amsmath}
\usepackage{amssymb}
\usepackage{setspace}
\usepackage{times}
\usepackage{helvet}
\usepackage{courier}
\usepackage{colortbl}
\newcolumntype{x}[1]{
>{\centering\hspace{0pt}}p{#1}}
\begin{document}

\title{Citing for High Impact}

\numberofauthors{3}

\author{
\alignauthor
Xiaolin Shi\\
       \affaddr{Stanford University}\\
       \affaddr{Stanford, CA 94305}\\
       \email{shixl@stanford.edu}
\alignauthor
Jure Leskovec\\
       \affaddr{Stanford University}\\
       \affaddr{Stanford, CA 94305}\\
       \email{jure@cs.stanford.edu}
\alignauthor
Daniel A. McFarland\\
       \affaddr{Stanford University}\\
       \affaddr{Stanford, CA 94305}\\
       \email{dmcfarla@stanford.edu}
}

\maketitle
\begin{abstract}
The question of citation behavior has always intrigued scientists from various disciplines. While general citation patterns have been widely studied in the literature we develop the notion of citation projection graphs by investigating the citations among the publications that a given paper cites. We investigate how patterns of citations vary between various scientific disciplines and how such patterns reflect the scientific impact of the paper. We find that idiosyncratic citation patterns are characteristic for low impact papers; while narrow, discipline-focused citation patterns are common for medium impact papers. Our results show that crossing-community, or bridging citation patters are high risk and high reward since such patterns are characteristic for both low and high impact papers. Last, we observe that recently citation networks are trending toward more bridging and interdisciplinary forms.
\end{abstract}




\section{Introduction}

\noindent When writing a research article, authors usually (though not always~\cite{simkin2005smc}) read and cite research they regard salient to the topic and approach being presented.  While it is true that multiple factors determine which references get made (e.g., disciplinary norms, strategic placement, etc.~\cite{adler-2009}), it is undeniable that the way citations are used can influence how well an article is received~\cite{Latour1988}. The selection of references situates a paper in a broader community of research and shapes readers' perception of it. Some articles cite narrowly in a single vein of well-defined research, others cite widely and idiosyncratically, and yet others span multiple coherent veins of work. The question this raises is how the pattern in which citations are presented influences how well an article is received.

To a great extent the patterns of citations may be a function of discipline. There are distinct norms of referencing within the hard sciences and social sciences.
As such, these norms represent distinct contexts of citing. In effect, the effectiveness of citation strategies likely depends on the scholarly domain in which the article is situated. As we show in this paper, each strategy of citation (specific to an article) has different returns on citation impact. An article whose citations span research communities may have the greatest returns of recognition in a field where segmented communities exist (juxtaposition with context). Or it may be that segmented communities require articles to reference only the relevant narrow community (alignment with context).

The field of scientometrics has long studied the impact of publications by applying bibliometrics to citation networks~\cite{dieks1976dis, redner-2005}. As early as in the 1960s, de Solla Price first developed models to explain the heavy tailed distribution in the citations an individual publication receives~\cite{desollaprice1965nsp}. The recent emergence of large-scale citation data has enabled the study of information flows between different areas in science~\cite{Bollen2009, boyack2005mbs}.
There are previous studies that show how different citation choices correspond to different citation impact \cite{Shi2009, Shi2009ICWSM}. The focus of the citation features of these studies is on the global citation network distance between the disciplines of the cited and citing papers.

In many social networks, the performance of individuals is tied to their local positions in their social networks~\cite{Burt_SocialCapital}. For example, there is a longstanding debate about the optimal network structure for the individuals' performance~\cite{lambiotte2009}. On the one hand, being clustered within tightly-knit communities benefits from fostering trust, facilitating the enforcement of social norms and common culture~\cite{Coleman1988, lambiotte2009}. On the other hand, networks with rich structural holes and weak ties are able to access heterogenous ideas and information more easily~\cite{Burt_SocialCapital, Granovetter82thestrength}. These features give an advantage to individuals who can connect distant parts of the network and get access to more diverse information and ideas. Such notions that the network position of a node is an indicator of the node's quality or performance also extend to web graphs and information networks. For example, algorithms for finding high quality web search results, like PageRank~\cite{PageRank} or Hubs and Authorities~\cite{Kleinberg1999}, rely heavily on the structure of the underlying web graph and the ``location'' of the node in this giant citation network of webpages. Moreover, the local structure of the web graph helps determine the overall quality of web search results and even predict users' searching behaviors~\cite{Leskovec2007}.

In this work, we build on the above intuitions, which were developed in the context of a web graph, and extend them to citation networks. We focus on the local graph structure of the citation network by investigating how patterns of citations vary between the scientific disciplines and how such patterns reflect the impact of the paper. We investigate the citations among the papers a chosen paper cites by defining {\em citation projection graphs}. Citation projection graphs not only examine the papers a paper refers to but also the citations among the referred papers. We use the graphical features of the citation subgraphs induced by cited publications and examine how scholarly publications draw previous information and knowledge together in different areas, as well as how this behavior correlates with the subsequent impact of the publications.

Our main finding is that there are significant differences in how high, low and medium impact papers position their citations. Whereas medium impact papers tend to cluster their citations in a narrow, well defined and connected field, the citation networks of both low and high impact publications cite a very diverse set of sources and refer to publications in various scientific communities. In this respect, high and low impact publications are similar. However, publications that are able to find bridges and connections between these scientific communities tend to be high impact, while publications that cite more idiosyncratically tend to be low impact. Our study indicates the high risk and high return of citing across communities in natural science and social science. We also find that disciplinary contexts and citation norms differ in different areas. For example, computer science behaves differently, as high impact papers there tend to have very focused citation networks mostly referring to papers from a narrow community. However, analyses of temporal trends of citation behaviors reveal that both natural and computer science are getting more interdisciplinary over time as the citation networks are getting more diverse and papers from multiple communities are cited.

The rest of the paper is organized as follows. In Section~\ref{sec:preliminaries}, we define the problem, introduce \emph{citation projection graphs}, and give a description of the data sets we use. In Section~\ref{sec:areas}, we examine the properties of the citation projection graphs in three different major areas: computer science, natural science, and social science. In Section~\ref{sec:randomgraphs},  the properties of citation projection graphs are compared with random graphs of same degree sequences. In Section~\ref{sec:diffimpact}, we study the correlations between the features of citation projection graphs and the impact of publications. Finally, the properties of citation projection graphs changing over time are studied in Section~\ref{sec:time}.  In Section~\ref{sec:conclusion}, we conclude our work, and discuss the findings and directions for future work.

\section{Approach and Datasets}\label{sec:preliminaries}

Next we present the methodology of citation projection graphs that are employed in the study and describe the citation network datasets.

\subsection{Citation networks}

Citation networks are networks of documents and the references among them. In this work, we focus on citation networks of scholarly publications, in which the nodes are publications and the directed edges are links from the citing to the cited papers. Information flows can then be interpreted as the scientific ideas and knowledge transmitted from one publication to another indicted by citation relationships. Two significant features of citation networks make them different from most other social or information networks: first, citation networks are directed graphs that are almost acyclic (this stems from the simple fact that only in rare case does a published work cite a future publication); second, in the evolution of citation networks, only new nodes and edges are added, while none are removed~\cite{Leicht2007,LeskovecDensification}. These two features guarantee that the subgraph induced by cited publications is fixed when a piece of work gets published.

\subsection{Citation projection graphs}
\label{sec:cpg}

\begin{figure}[t]
  \centering
  \includegraphics[width=0.6\columnwidth]{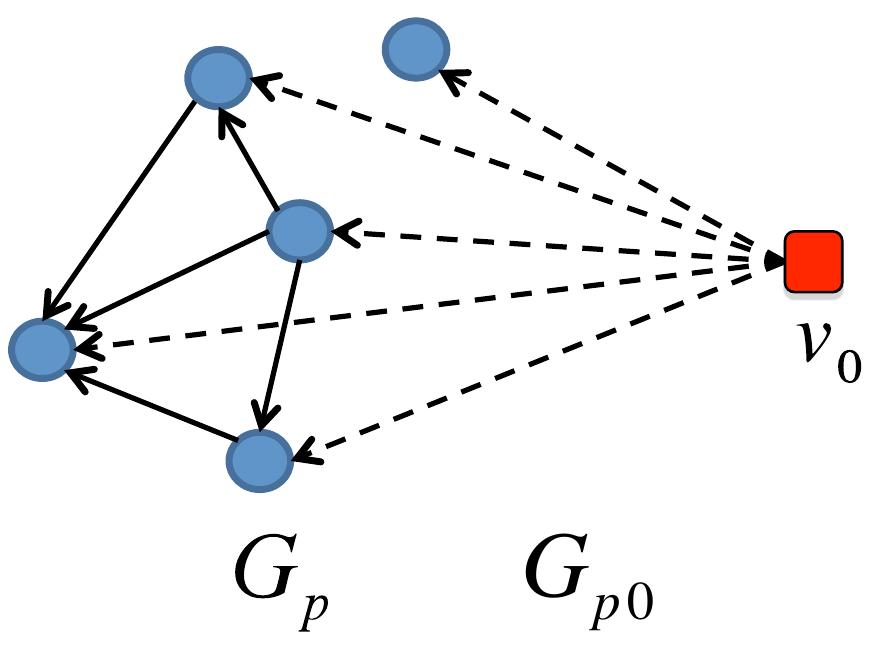}
  \caption{The citation project graphs $G_p$ and $G_{p0}$ of publication $v_0$. The red square node is the publication $v_0$, and
  the blue circle nodes represent publications cited by $v_0$. $G_p$ is the induced subgraph on the blue circle nodes (solid edges) and
  represents references between the papers cited by $v_0$. $G_{p0}$ is composed of $G_p$ together with the red square node $v_0$ and the
  dashed edges.}
  \label{fig:CitedExample}
\vspace{-2mm}
\end{figure}

In order to study the features of the citation behavior of publications, we study the properties of publication {\em citation projection graphs}. Intuitively, we take all the papers a given article cites, ``project'' them on the underlying citation graph and then extract a subgraph of citations among all the cited papers. This way we are able to capture both the cited papers and the relationships (i.e., citations) among them. More precisely, we define two types of citation projection graphs associated with a publication $v_0$. First is the subgraph $G_p$ that is induced by the papers cited by $v_0$, and the second is the subgraph $G_{p0}$ that is induced by the cited papers together with $v_0$ itself. Figure~\ref{fig:CitedExample} shows an example of a citation projection graph of a paper $v_0$ represented by a square node. We take all the references of $v_0$ represented by circular nodes and extract them from the citation graph together with the citations among the circular nodes (represented by bold directed links).

\begin{figure}[t]
  \centering
  \includegraphics[width=0.75\columnwidth]{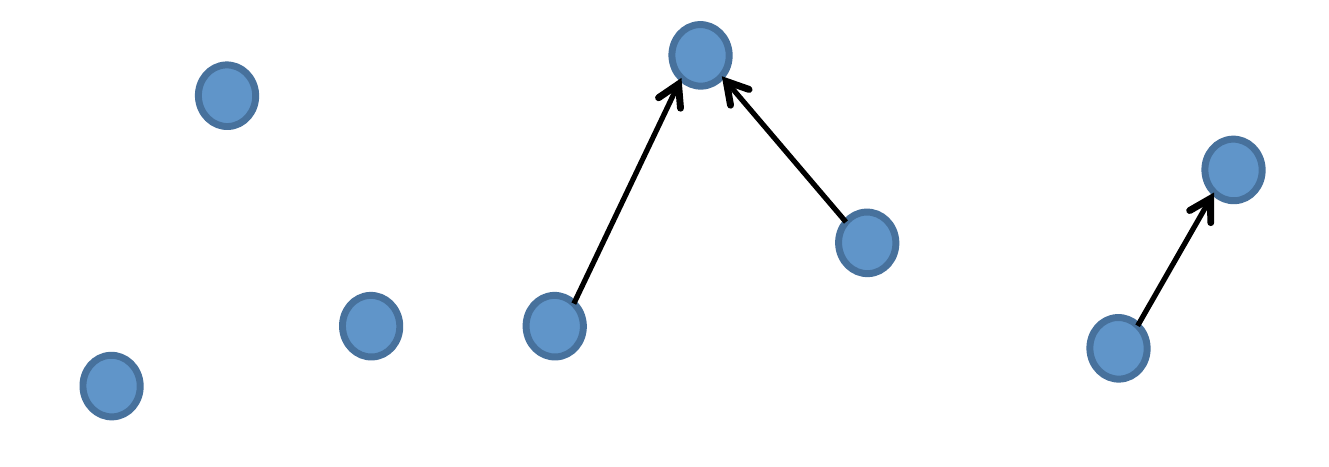} \\
  (a) An idiosyncratic citer \\
  \includegraphics[width=0.68\columnwidth]{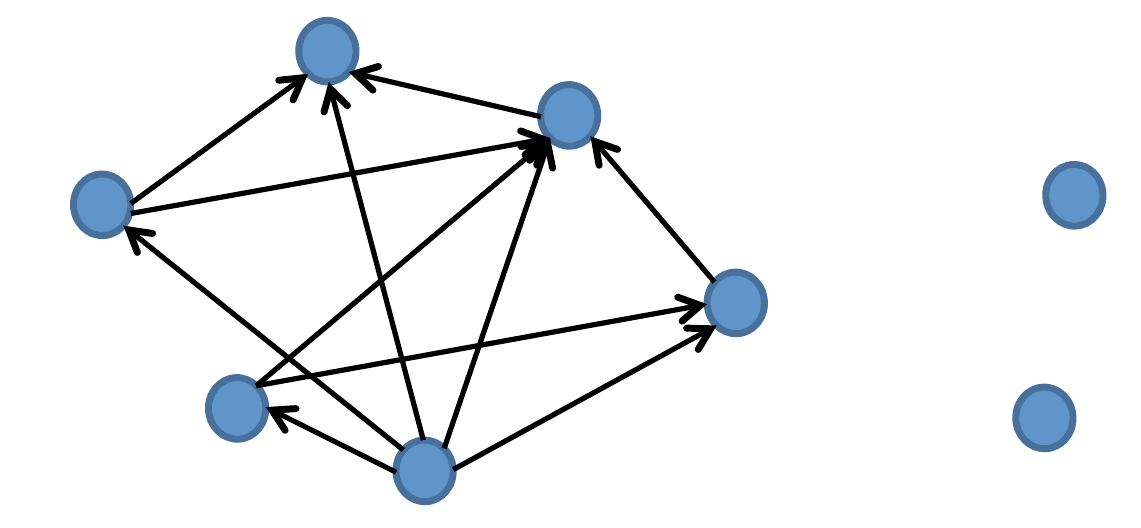} \\
  (b) A within-community citer \\
  \includegraphics[width=0.78\columnwidth]{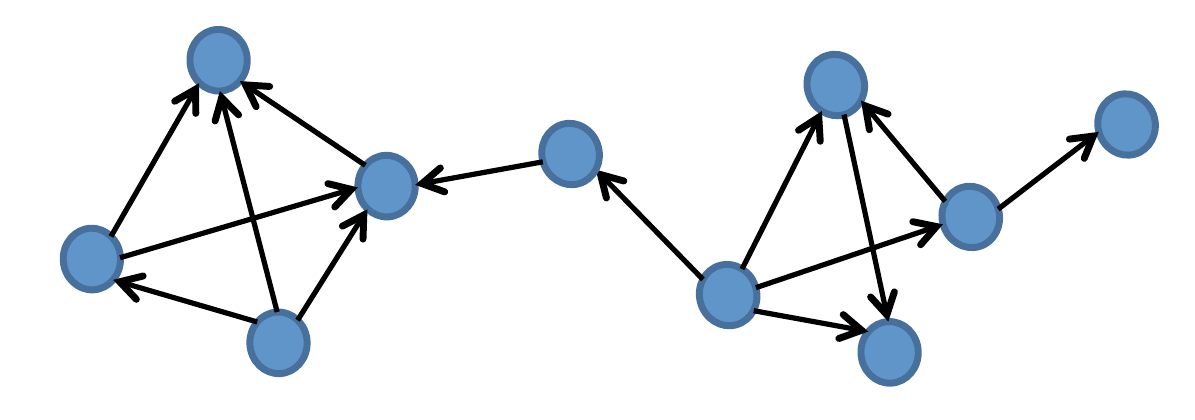} \\
  (c) A brokerage citer \\
  \caption{The schematic representation of three prototypical types
  of citation projection graphs $G_p$.}
  \label{fig:examples}
\vspace{-2mm}
\end{figure}

Using the projections graphs we then define a set of network metrics that describe and characterize the structure of the network created by the references between papers that $v_0$ refers to. To give some idea about what classes of projection graphs we aim to distinguish, we show in Figure~\ref{fig:examples} schematic representations of three prototypical classes of citation projection networks. Figure~\ref{fig:examples}(a) shows a citation projection network $G_p$ of an idiosyncratic paper that creates ``random'' references across various disciplines. This creates a citation projection that is sparsely connected. Figure~\ref{fig:examples}(b) illustrates a network of a within-community citer, where most of the references of a paper focus on a narrow fields or a set of papers that are very well connected among themselves. Thus we expect citation projection networks of such papers to contain a large, densely connected component with many citations among the papers. Figure~\ref{fig:examples}(c) shows an example of a brokerage citer, where different clusters of papers that intuitively correspond to different scientific disciplines or communities are cited but connections between these fields are also identified. Such projection networks are characterized by high betweenness nodes and medium link density.

Our aim now is to define a set of network metrics or statistics that will be able to compare and distinguish between the three prototypical classes of the above citation projection networks. For every publication $v_0$, we use six metrics (\textbf{M1}-\textbf{M6}) to characterize the features of the sub-graphs induced by the cited publications. The first four metrics are with regard to the projection graph $G_p$, and the last two are with regard to the projection graph $G_{p0}$:

\begin{itemize}
  \item {\bf (M1) Graph density}: $D = \frac{2|E|}{|V|(|V|-1)}$, where $V$ are the set of nodes and $E$ are the set of
      edges in the citation projection subgraph $G_p$. Intuitively, we expect papers that cite within-community
      (Figure~\ref{fig:examples}(b)) to have densest projection networks, while idiosyncratic citers
      (Figure~\ref{fig:examples}(a)) would have the lowest.
  \item {\bf (M2) Clustering coefficient}: the average fraction of closed triangles between the connected triples of nodes in $G_p$. Average clustering coefficient $C=\frac{1}{|V|} \sum_{v_i} C(v_i)$, where 
      for every node $v_i \in V$ we compute:
      $$
      C(v_i) = \frac{\textrm{number of closed triads connected to $v_i$}}{\textrm{number of triples of vertices centered
      on $v_i$}}
      $$
      This metric captures the degree of cohesiveness in the network and aims to identify citation projection networks
      with many ``friend-of-a-friend'' connections. Denser networks will generally have higher clustering coefficient;
      however, the locality of the edges also plays an important role in the clustering coefficient.
  \item {\bf (M3) Connectivity}: fraction of nodes in the largest (weakly) connected component of $G_p$. We expect
      networks of highly focused papers with a highly specific and narrow set of references (Figure~\ref{fig:examples}(b))
      to have a relatively large maximum connected component. Similarly, papers that cite across disciplines but find
      connections between them (Figure~\ref{fig:examples}(c)) also have a large connected component.
  \item {\bf (M4) Maximum betweenness}: the highest betweenness of nodes in $G_p$. The betweenness of a node
      $v_i \in V$ in $G_p$ is the centrality measure of the fraction of all shortest paths that pass through $v_i$ in the graph:
      $$
      B(v_i) =  \sum_{j<k} g_{jk}(v_i)/g_{jk}
      $$
      where $g_{jk}$ is the number of shortest paths linking nodes $v_j$ and $v_k$, and $g_{jk}(v_i)$ is the subset of
      those paths that pass node $v_i$. Using maximum betweenness we will be able to separate the idiosyncratic and
      within-community citers from the brokerage citers. For brokerage citers the papers that link separate clusters
      will have high betweenness since they act as connectors between the disciplines, while idiosyncratic and
      within-community citers will not have high maximum betweenness for two reasons: for within-community citers
      there exist many different equivalent shortest paths between pairs of nodes and thus no node will have
      a particularly high betweenness; similarly, for idiosyncratic citers there are simply no edges to connect
      different nodes and thus maximum node betweenness will be naturally small.
  \item {\bf (M5) Betweenness of $v_0$}: the betweenness centrality of $v_0$ in the citation projection graph $G_{p0}$.
      This is computational the same as (M4) but now the graph is different as we also include node $v_0$ in the
      projection graphs. Intuitively, the betweenness of $v_0$ captures how good the connector $v_0$ itself is. For
      example, for networks like Figure~\ref{fig:examples}(a) betweenness of $v_0$ is very high as every circular node in
      the network also connects to $v_0$ and thus the majority of shortest paths will pass through $v_0$. On the contrary,
      for networks such as Figure~\ref{fig:examples}(b) and (c) the betweenness of $v_0$ will be lower since between
      many pairs of nodes in the network there already exist relatively short paths.
  \item {\bf (M6) Network constraint of $v_0$}: the network constraint of $v_0$ in $G_{p0}$ measures the extent to which
      an individual node's interaction with others is concentrated in a single group of interconnected neighbors
      \cite{burt1992}:
      $$
      NC(v_i)=\sum_{j \neq i}(\sum_{k \neq j}(p(i,k)p(k,j))^2)
      $$
      where $p(ij) = e(i, j)/\sum_{k}e(i, k)$, and $e(i, j)$ is the weight on the edge connecting $v_i$ and $v_j$. The
      network constraint varies with three network dimensions: size, density and hierarchy. A node that is well embedded
      into the citation network (for example as in Figure~\ref{fig:examples}(b)) will have high network constraint while
      nodes that act as bridges or brokers between fields and disciplines (Figure~\ref{fig:examples}(a),(c)) will have low
      network constraint.
\end{itemize}

To summarize, the six metrics we use in this study are able to capture the following three major aspects of paper citation projection networks: (1) The extent to which citations are focused and clustered regions of dense citation (\textbf{M1, M2}); (2) Random or idiosyncratic citations (\textbf{M3, M4}); (3) Citations that bridge groups (\textbf{M5, M6}).

\begin{figure*}[th]
  \centering
  \begin{tabular*}{\textwidth}{@{\extracolsep{\fill}}ccc}
    \includegraphics[width=0.667\columnwidth]{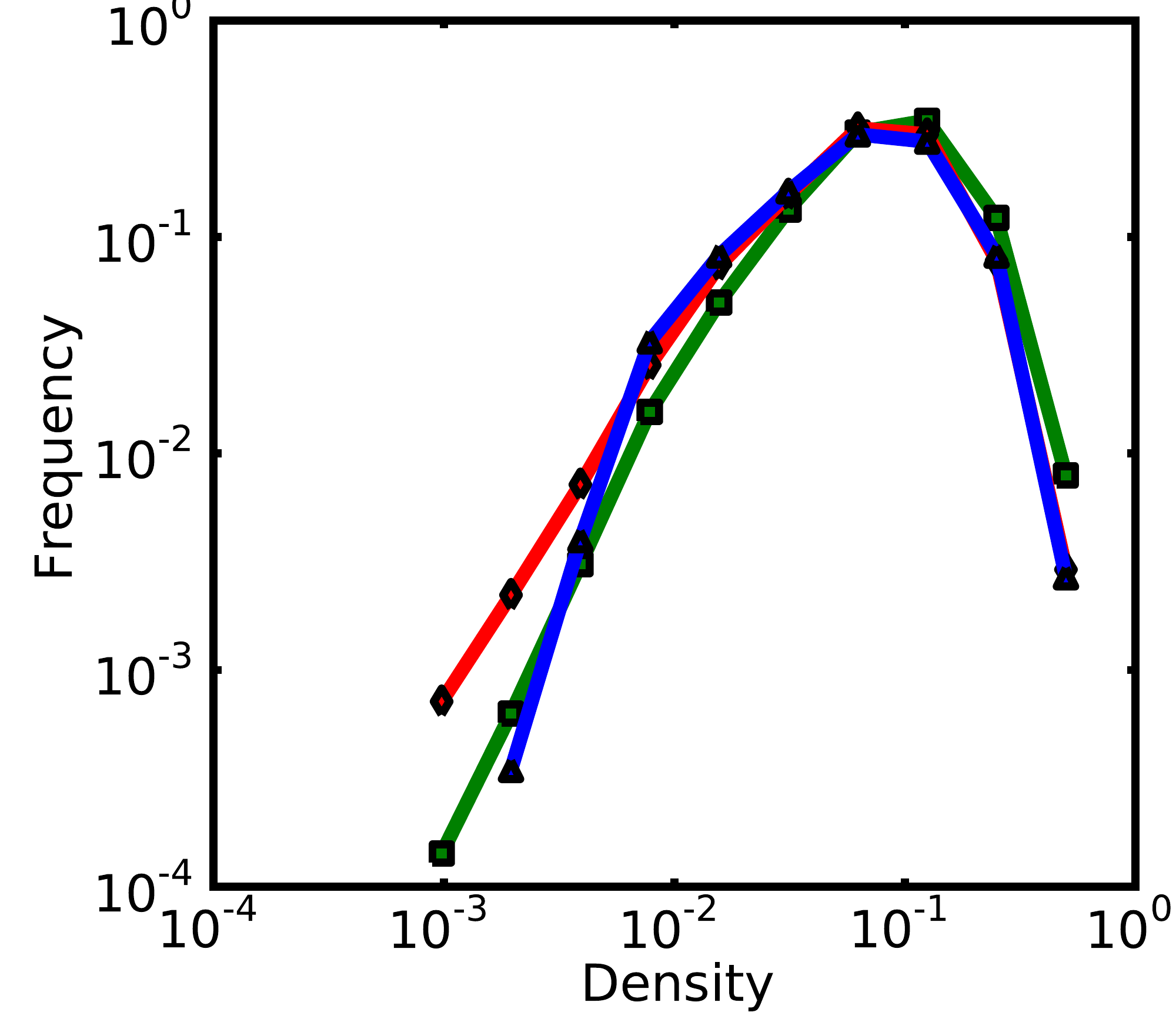} &
    \includegraphics[width=0.667\columnwidth]{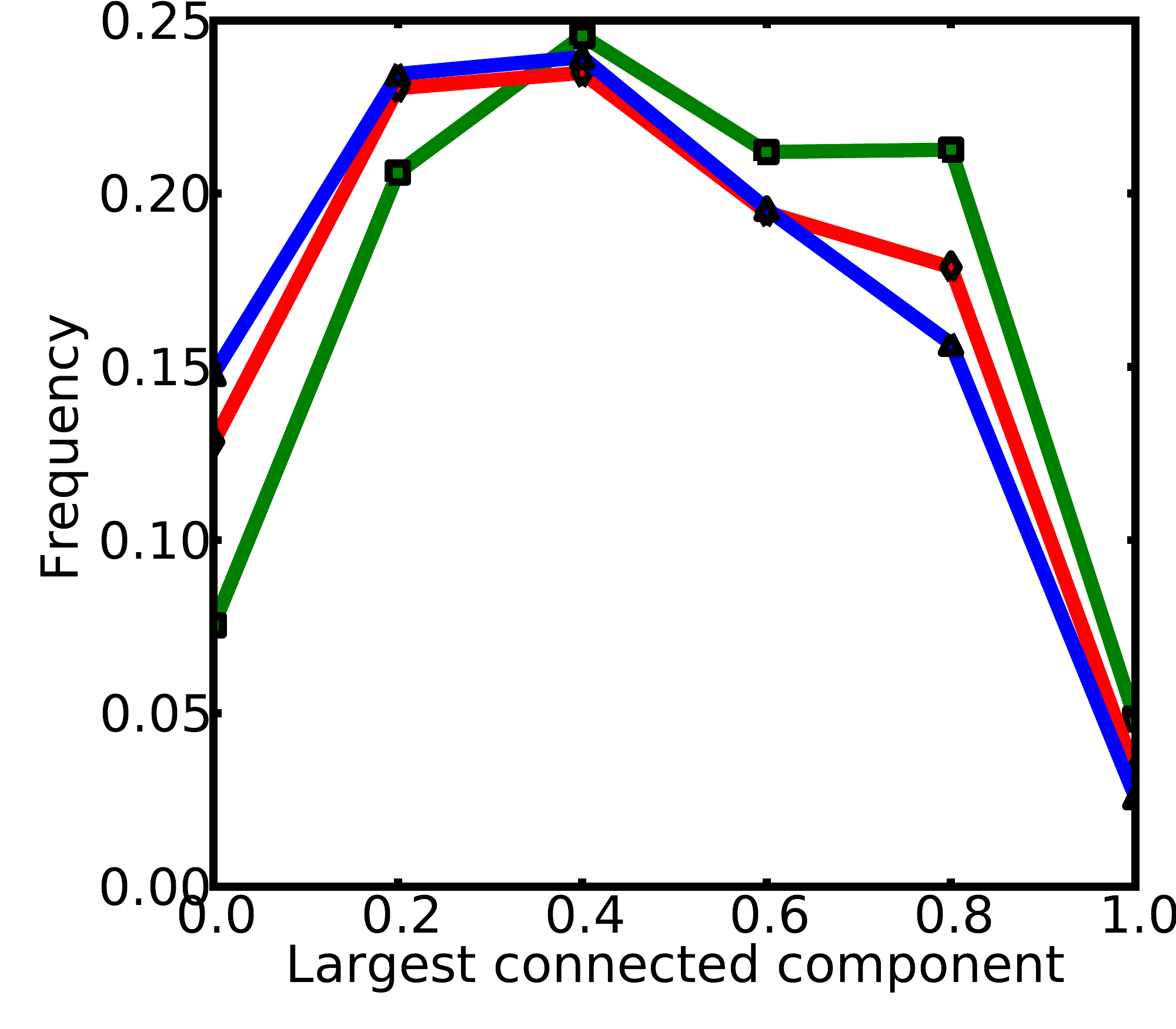} &
       \includegraphics[width=0.667\columnwidth]{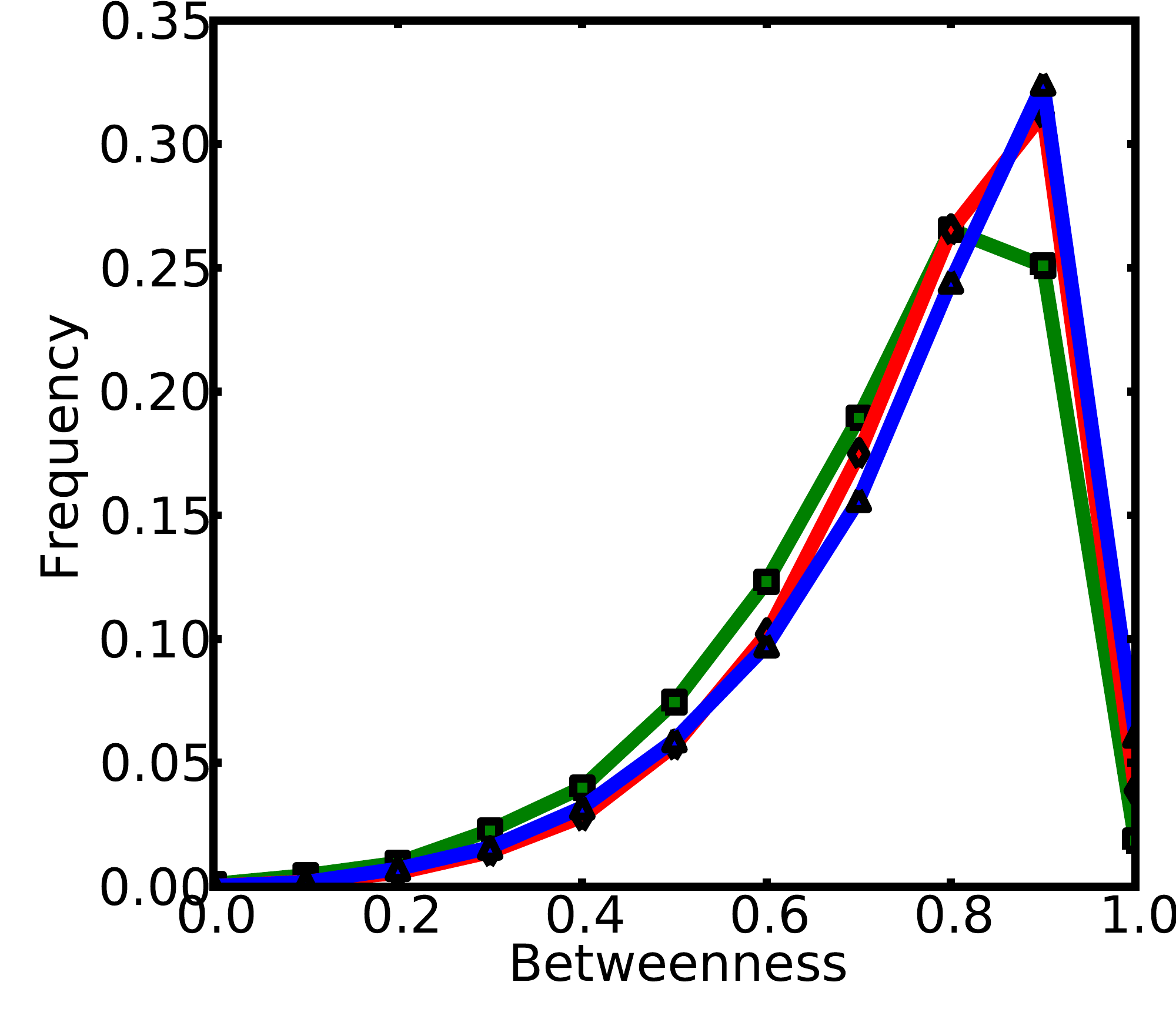}  \\
    (a) M1: Density & (c) M3: Connectivity & (e) M5: Betweenness \\
    \includegraphics[width=0.667\columnwidth]{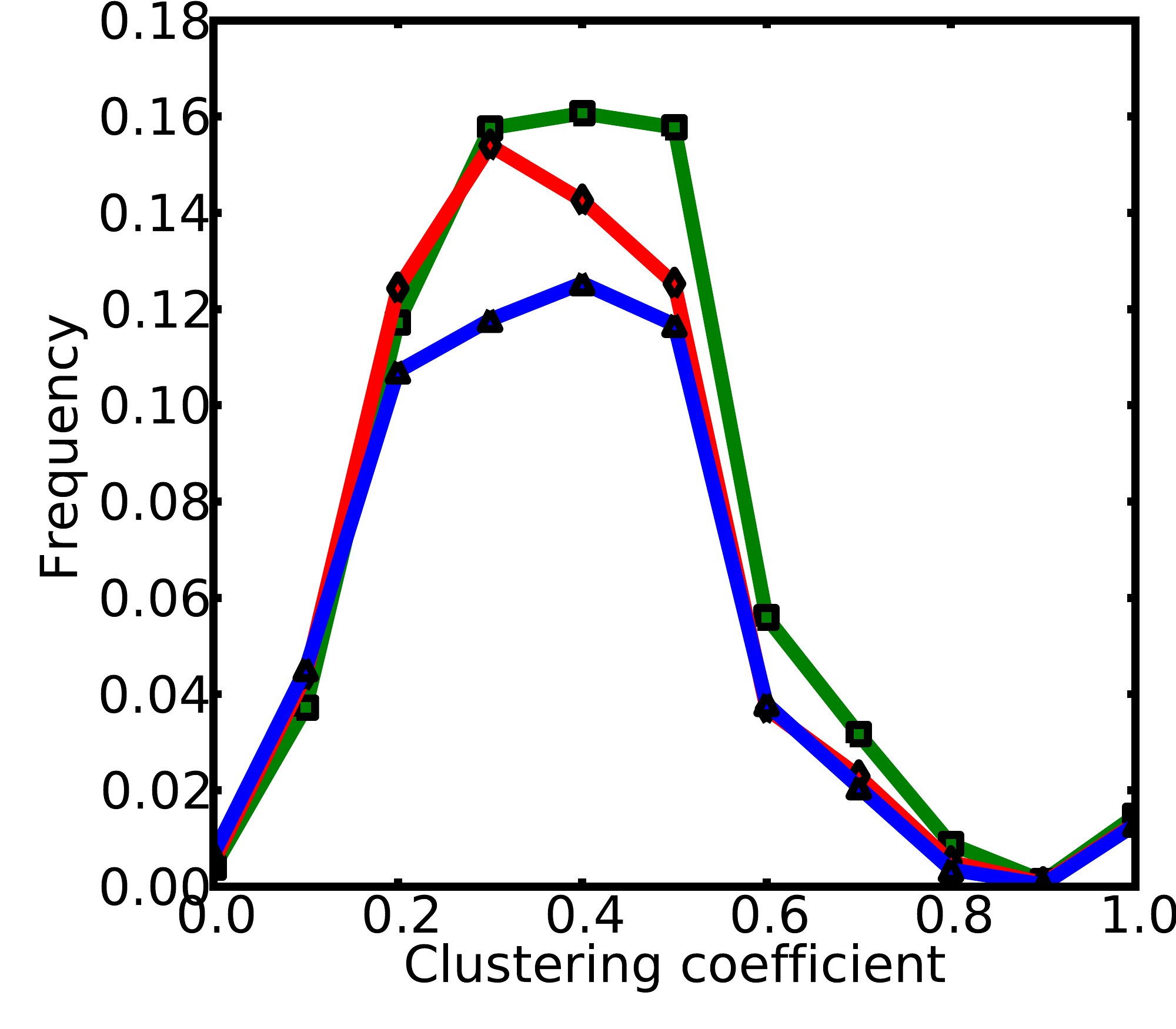} &
    \includegraphics[width=0.667\columnwidth]{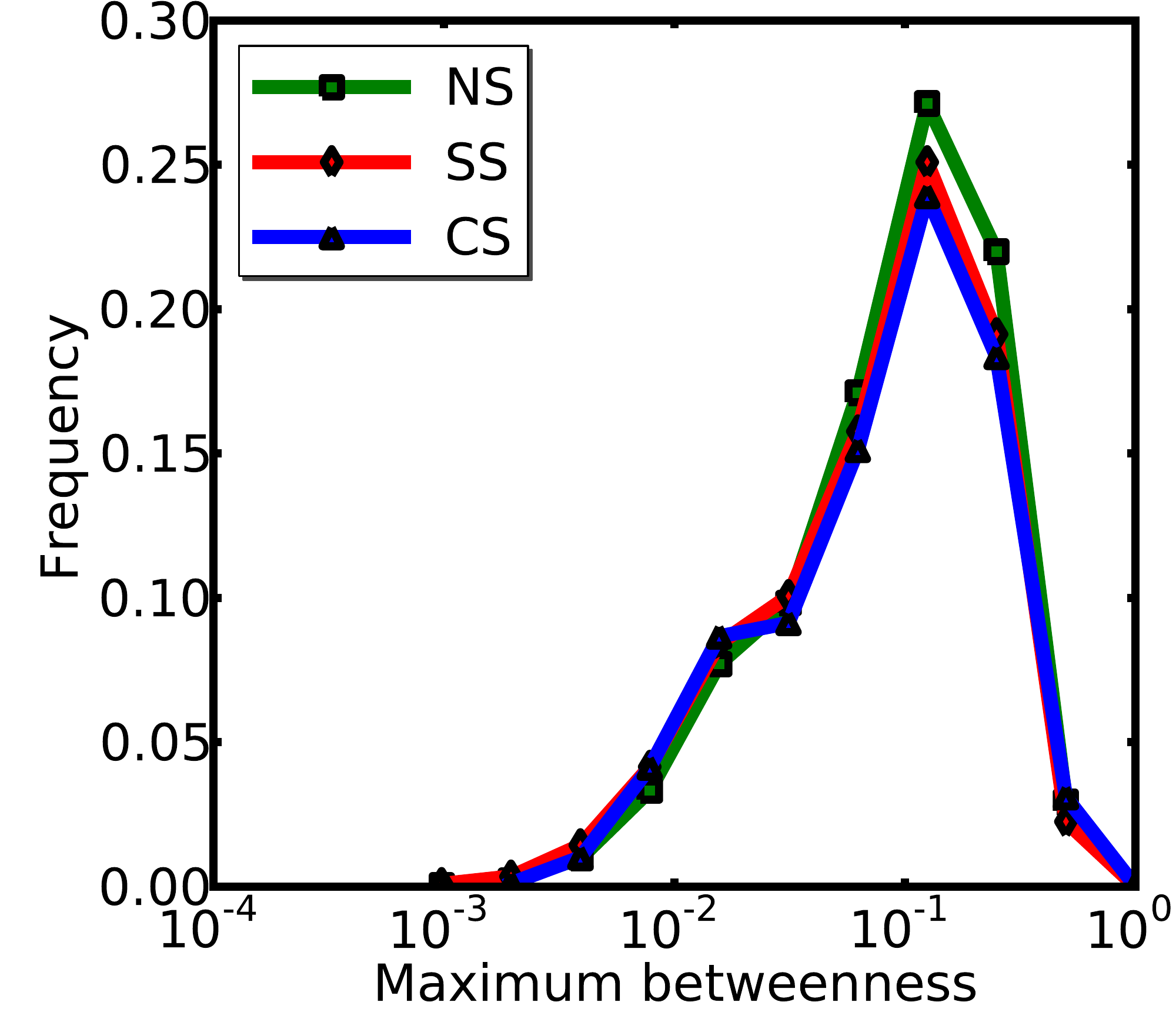} &
    \includegraphics[width=0.667\columnwidth]{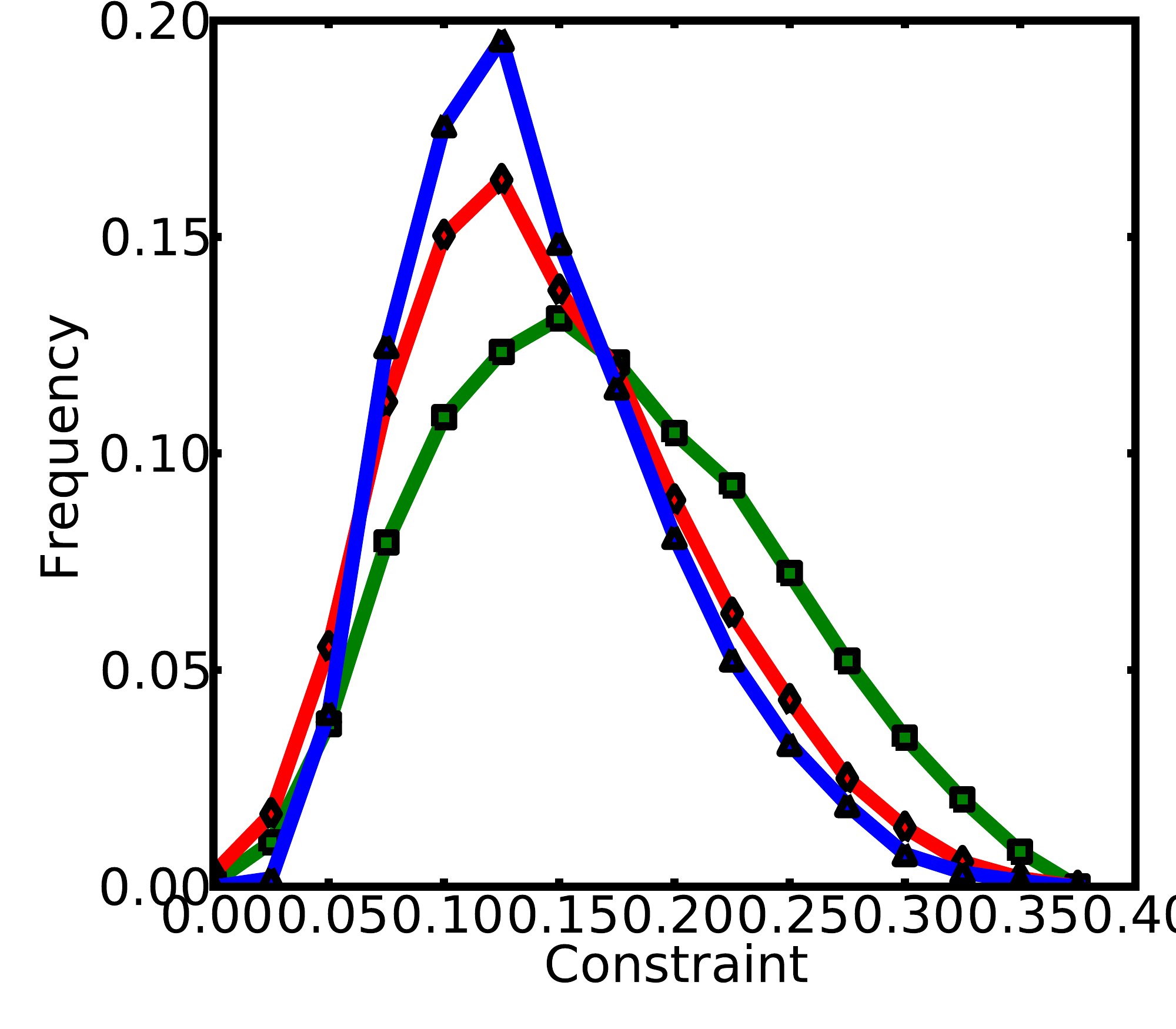}  \\
    (b) M2: Clustering Coefficient & (d) M4: Maximum Betweenness of $v_0$ & (f) M6: Network constraint of $v_0$ \\
  \end{tabular*}
    \caption{Properties of Citation Projection Graphs of publications in natural science (green squares), social science (red diamonds) and
    computer science (blue triangles). Natural science publications have more cohesive and focused citations networks than papers
    from computer science or social science.}
  \label{fig:areas}
\vspace{-2mm}
\end{figure*}

\subsection{Dataset description}\label{sec:datadescription}

We use the ACM (Association for Computing Machinery) and JSTOR (Journal Store) datasets in this study. In the ACM data set, there are 613,444 conference and journal papers, most of which concern computer science. There are about 346,000 citations among them. While computer science is a diverse field, the references in the ACM corpus are specific to the general domain of Computer Science and under-represents linkages across disciplines. For this reason, we also employ the JSTOR corpus which has references that span multiple disciplines. JSTOR has 878,841 research articles in 1108 journals, classified into 47 disciplines according to their venues, roughly corresponding to 3 sets: 435,425 in natural science, 290,703 in social science and 86,410 in arts \& humanities. The rest of 66,303 research articles are on the boundary (belong to more than one major sets).  There are 6,585,136 citations in total. These citations, limited to the cases where both the citing and cited articles are in the dataset, are a subset of the 23,451,235 citations made by all the articles.

In the following analyses, we compute the values of the properties of citation projection graphs of every publication that cited more than 10 other papers in the dataset. In this way, the publications that are not well represented in the data set are eliminated. After this screening, 49,290 research articles remain in natural science, 40,531 research articles in social science, and 11,565 articles in ACM. We do not analyze humanities articles in JSTOR because that subset of articles is very small and prone to measurement errors.

\section{Citation Projection Graphs in \\
different areas of science} \label{sec:areas}

\begin{figure*}[t]
  \centering
  \begin{tabular*}{\textwidth}{@{\extracolsep{\fill}}ccccc}
  \hline \hline
  \multicolumn{5}{c}{\textbf{Natural Science}}\\
  \hline \hline \\
%  \vspace{0.5mm}\\
   \includegraphics[width=0.37\columnwidth]{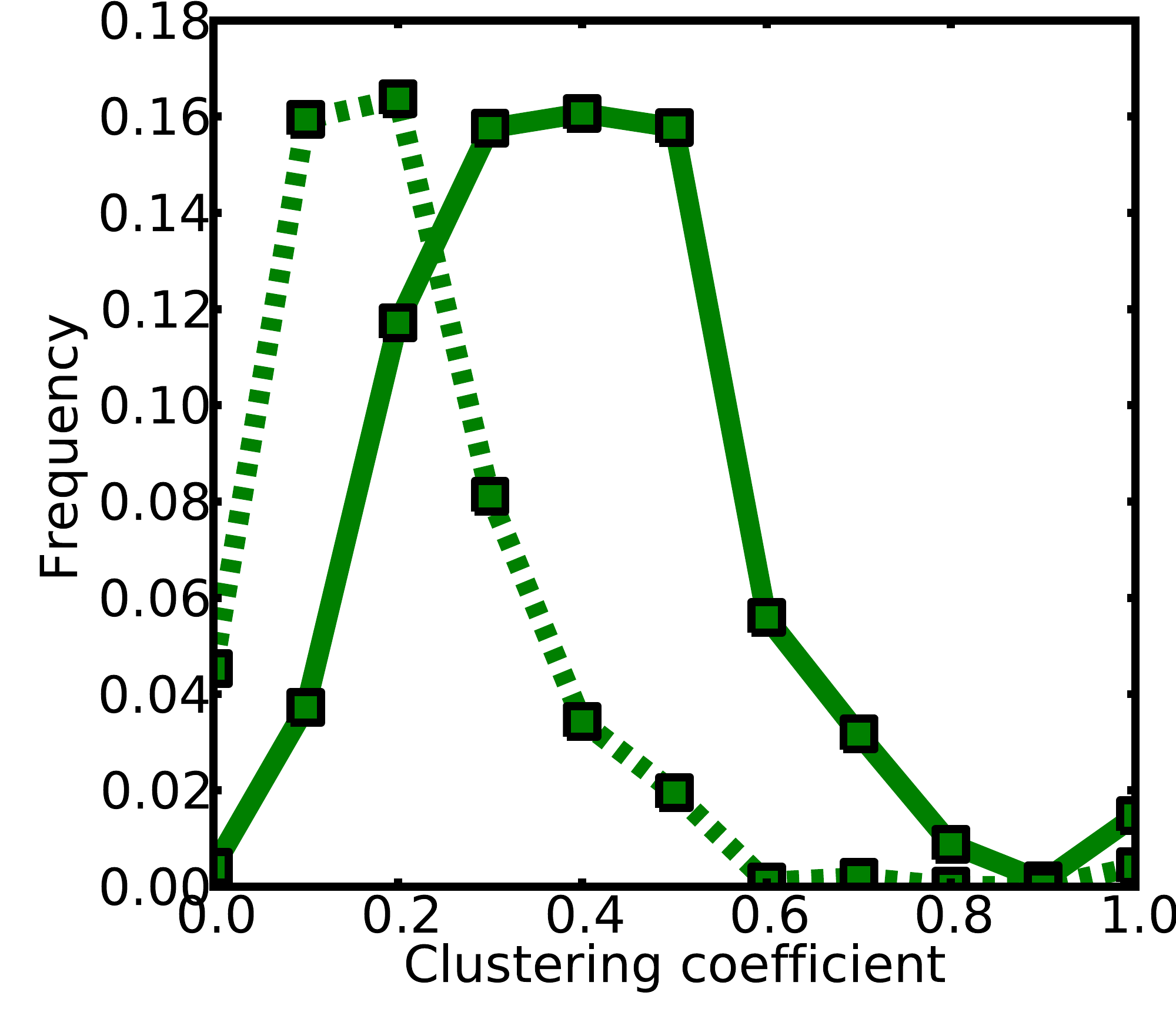} &
  \includegraphics[width=0.37\columnwidth]{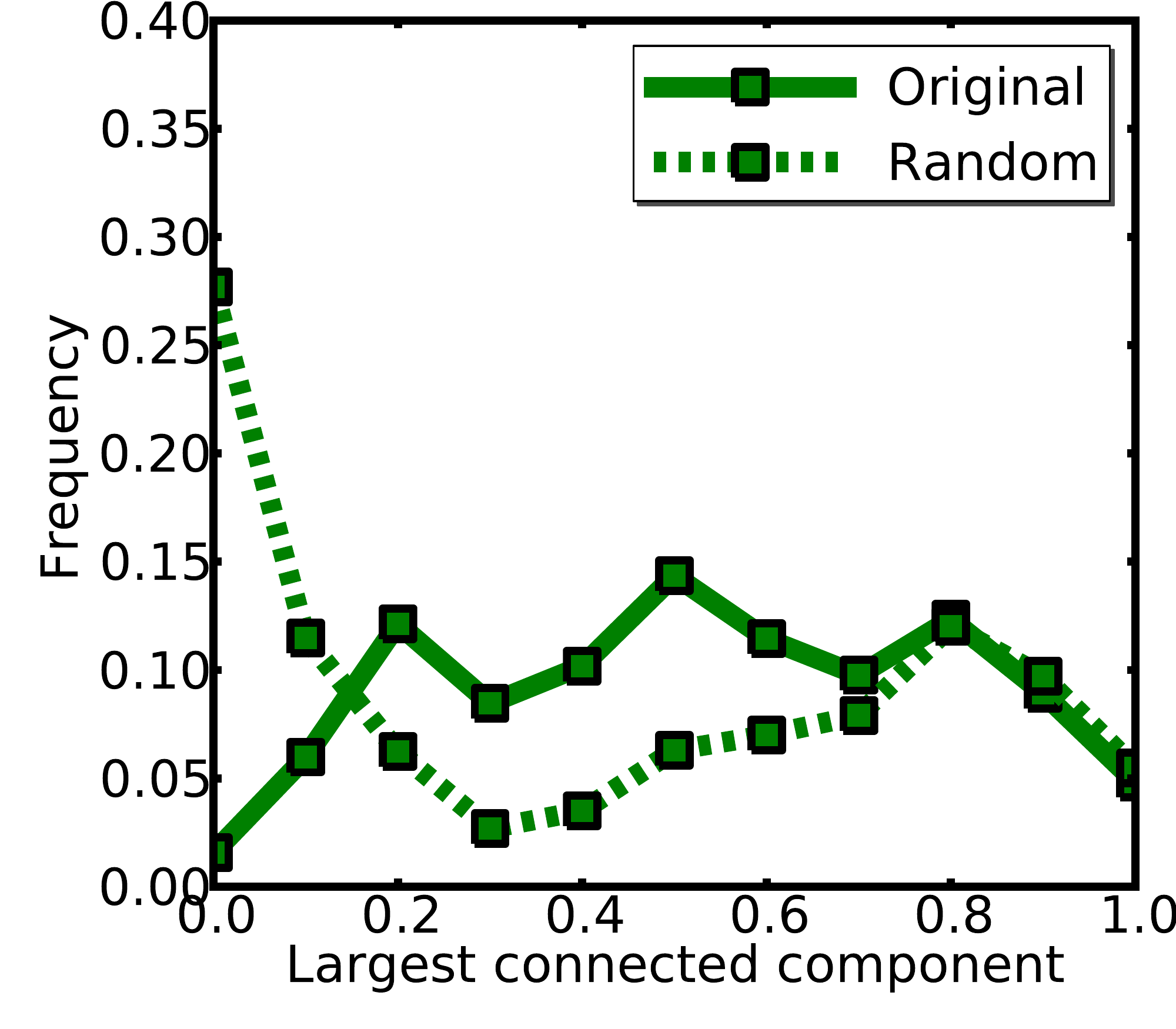} &
  \includegraphics[width=0.37\columnwidth]{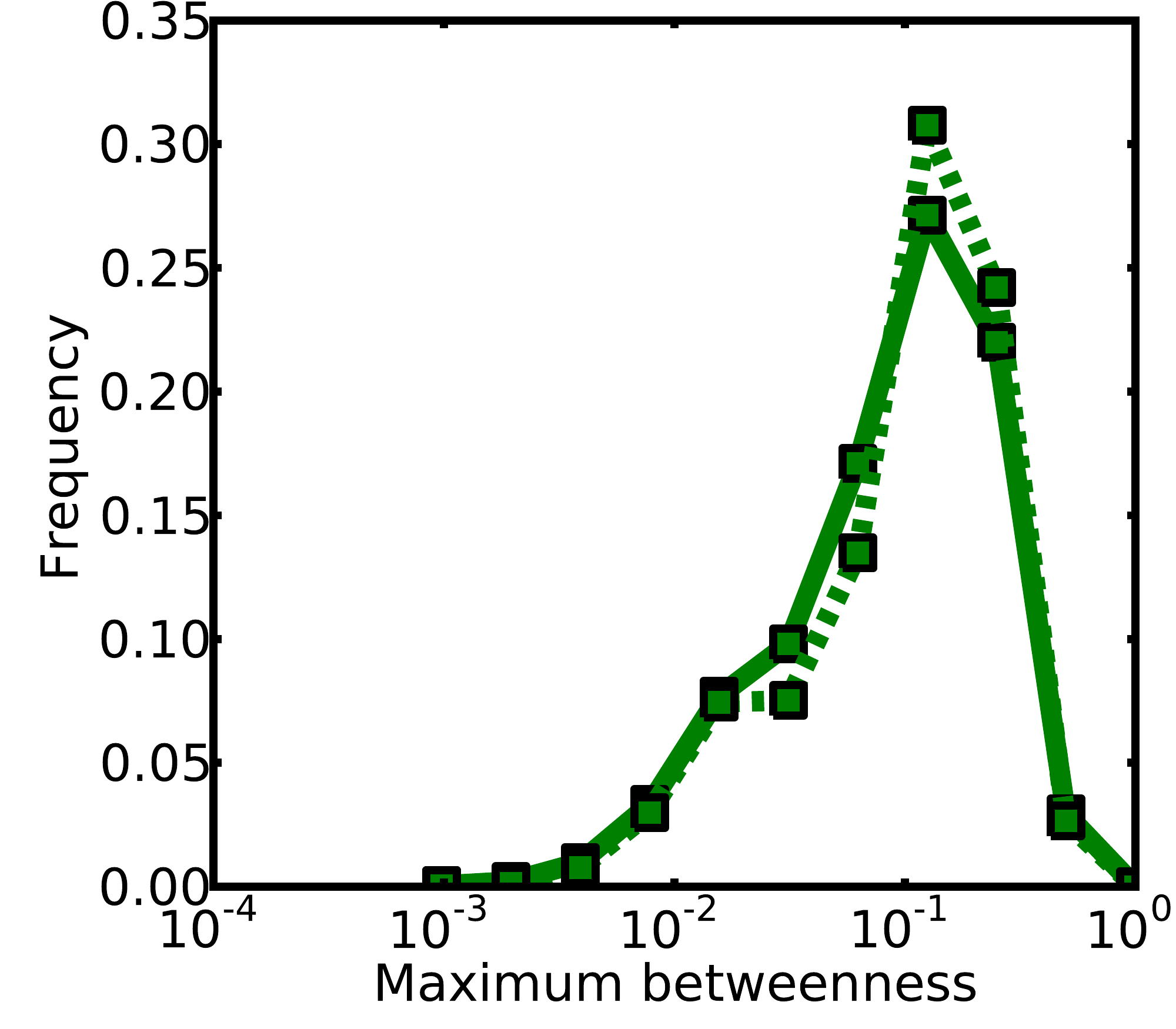} &
  \includegraphics[width=0.37\columnwidth]{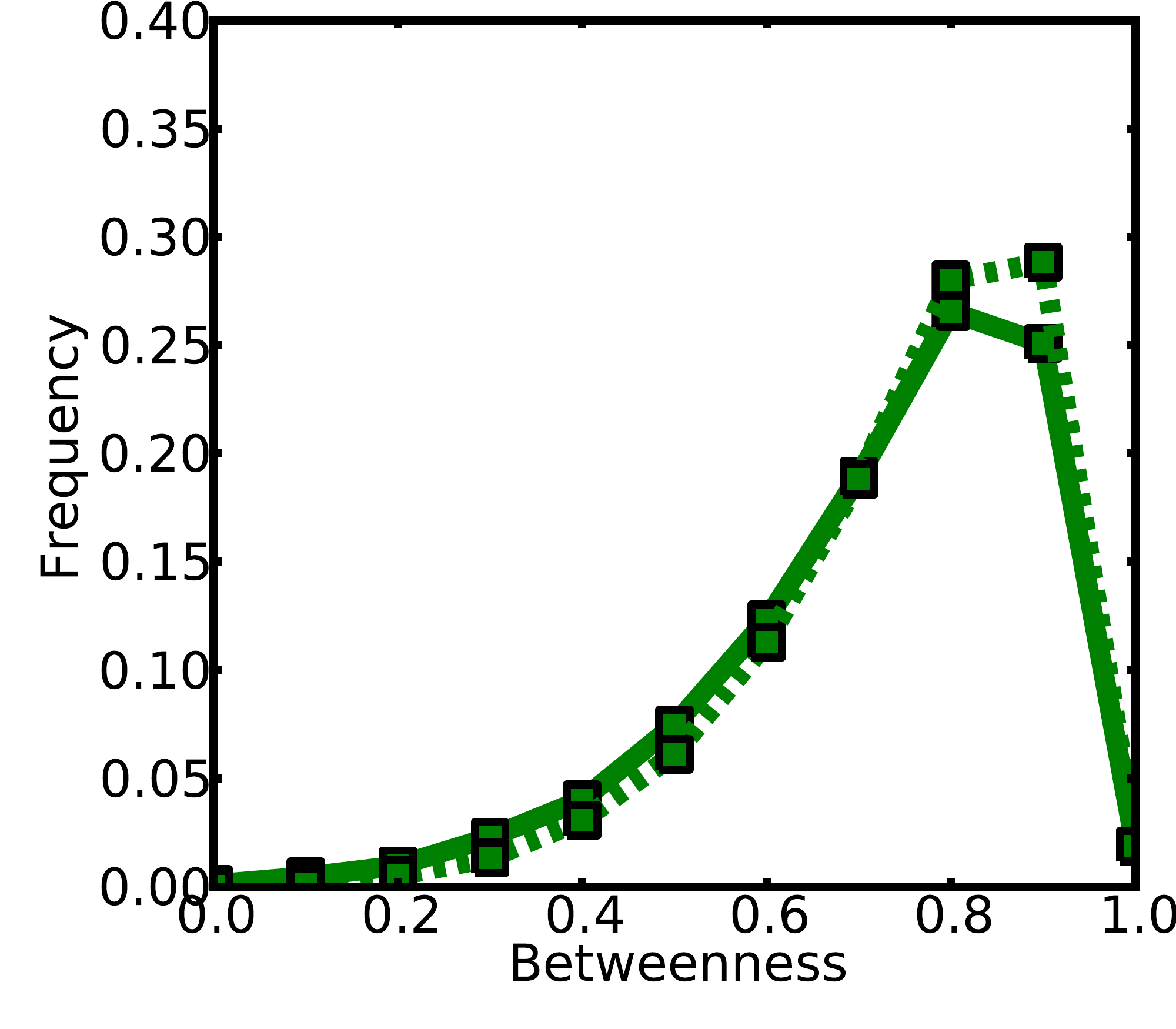} &
  \includegraphics[width=0.37\columnwidth]{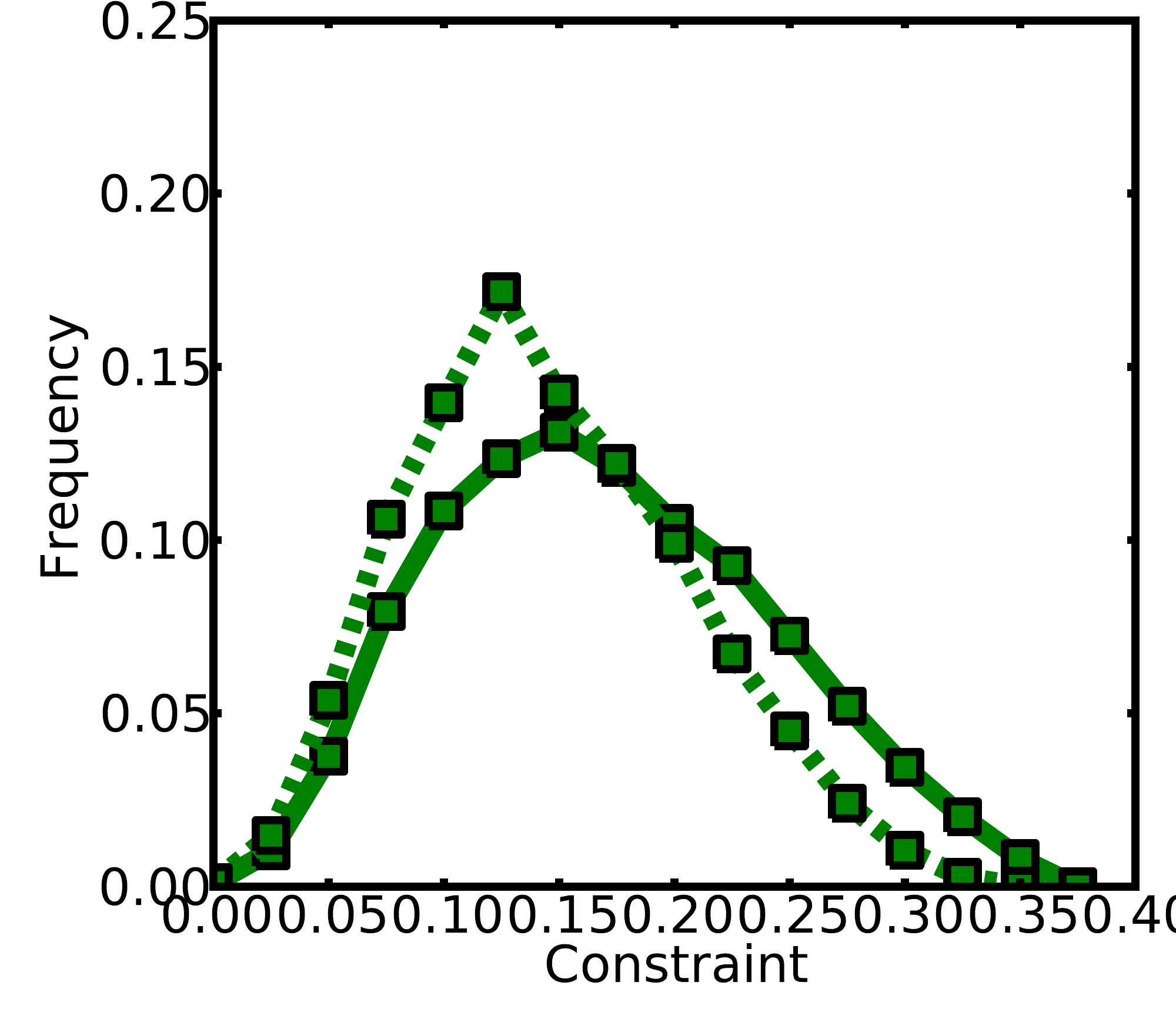} \\
  \hline \hline
  \multicolumn{5}{c}{\textbf{Social Science}}\\
  \hline \hline \\
   \includegraphics[width=0.37\columnwidth]{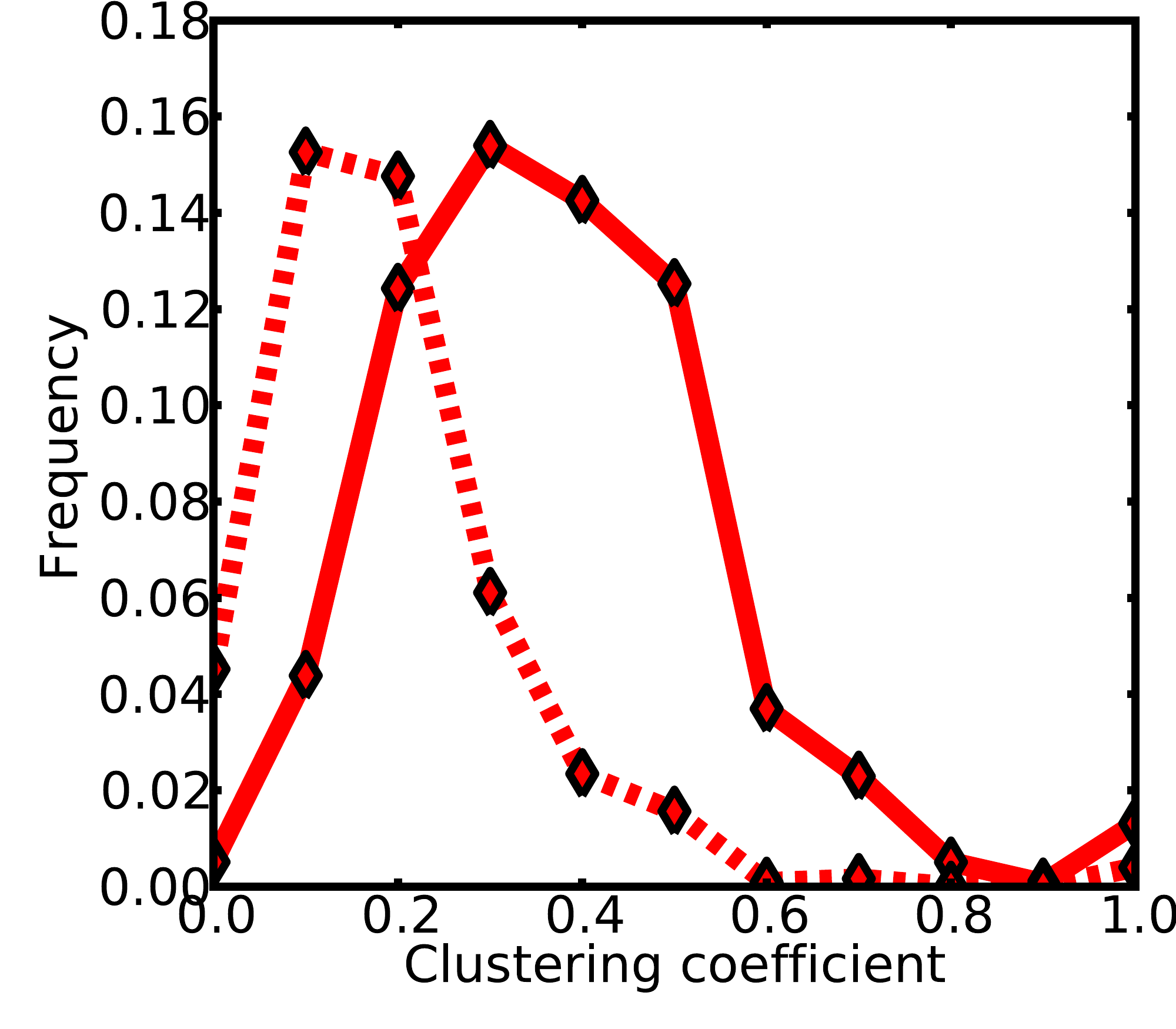} &
  \includegraphics[width=0.37\columnwidth]{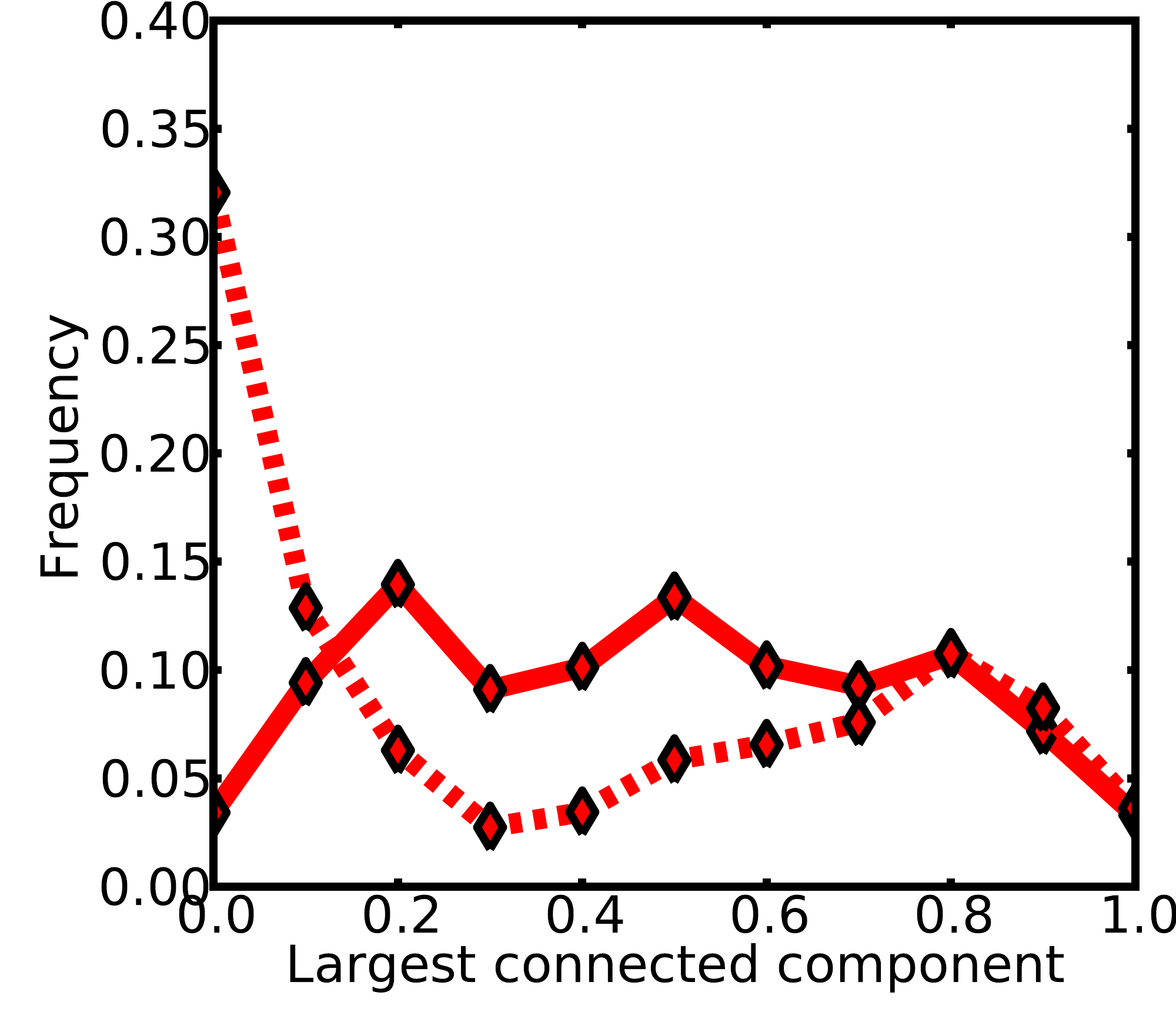} &
  \includegraphics[width=0.37\columnwidth]{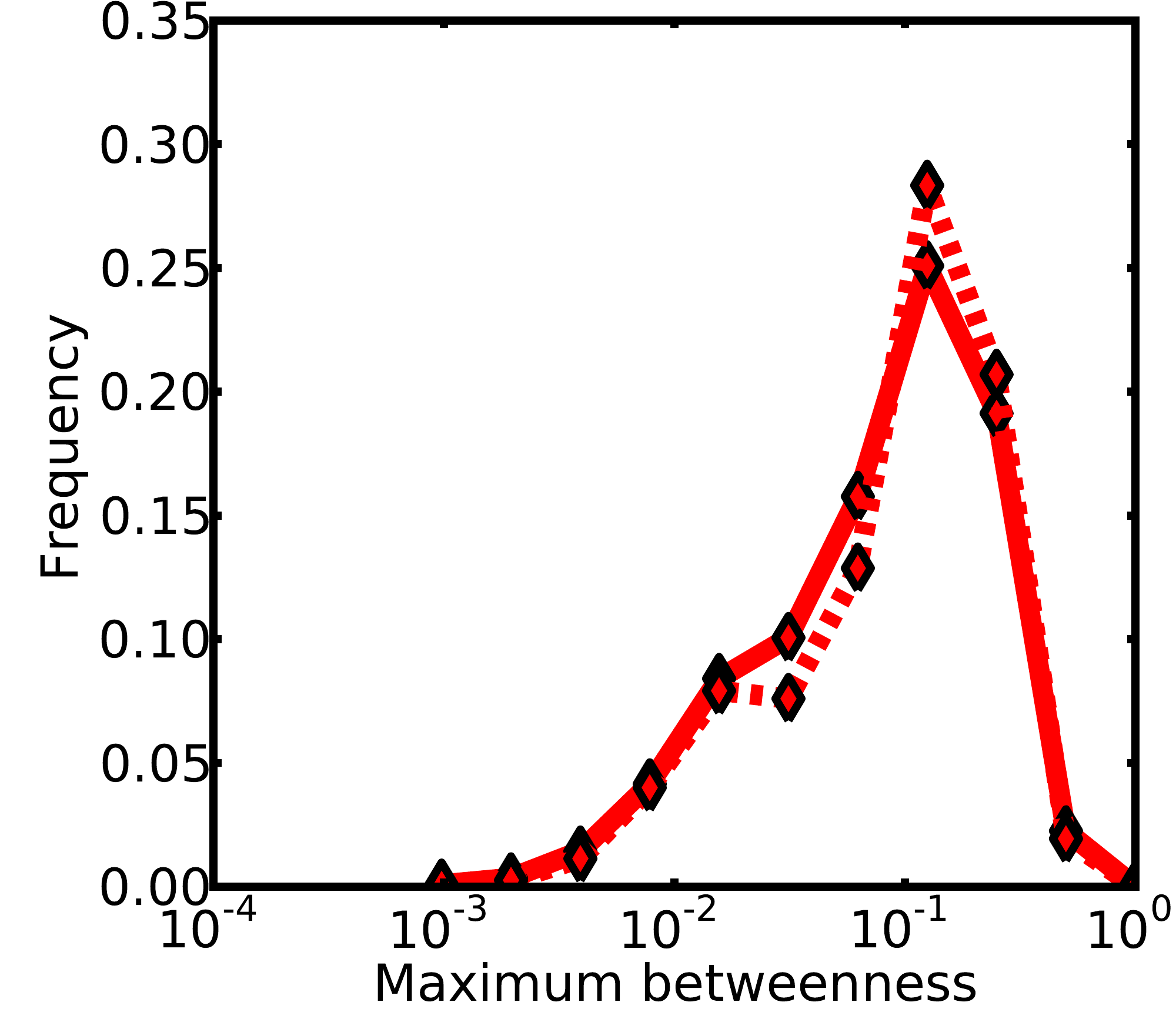} &
  \includegraphics[width=0.37\columnwidth]{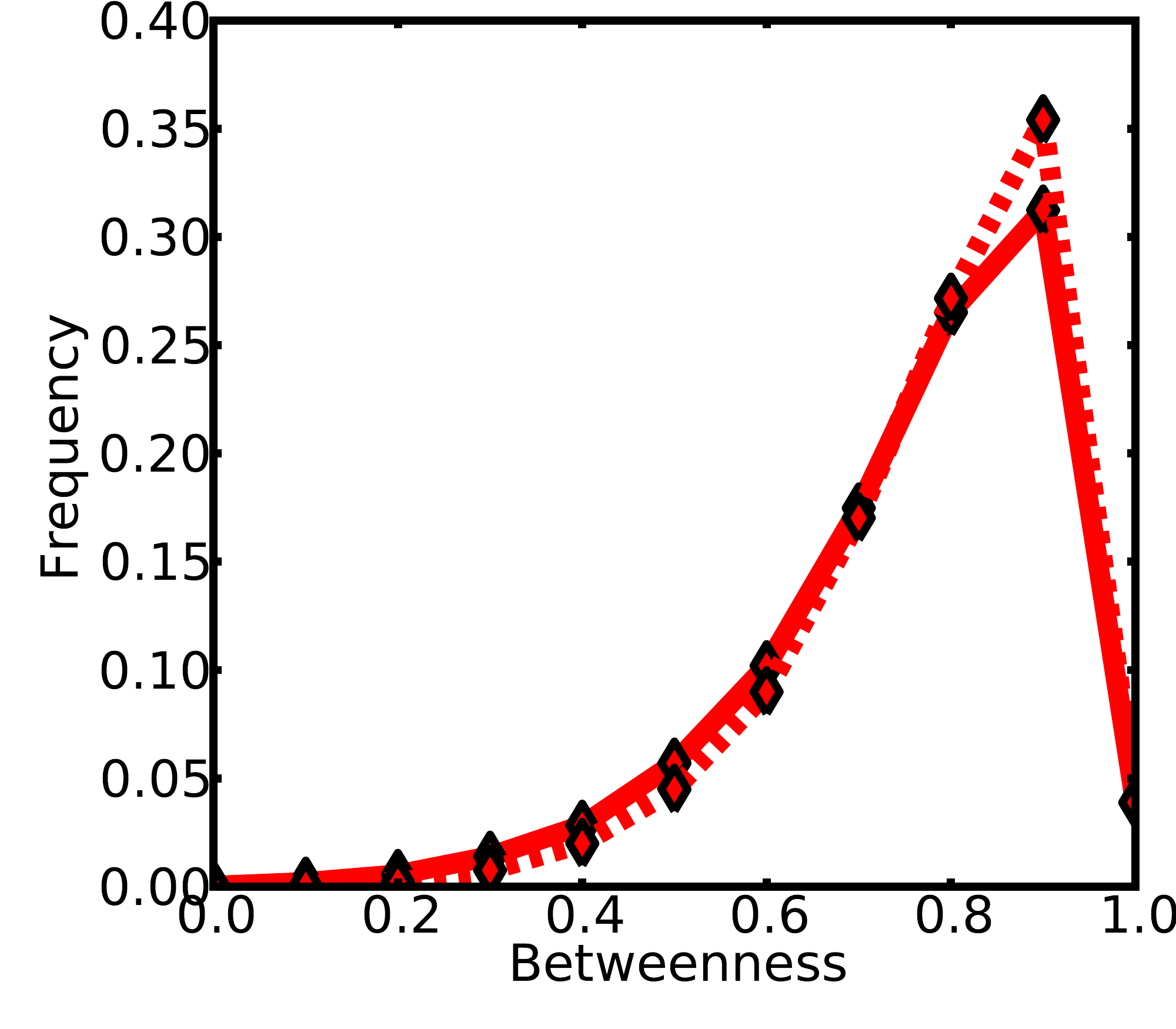} &
  \includegraphics[width=0.37\columnwidth]{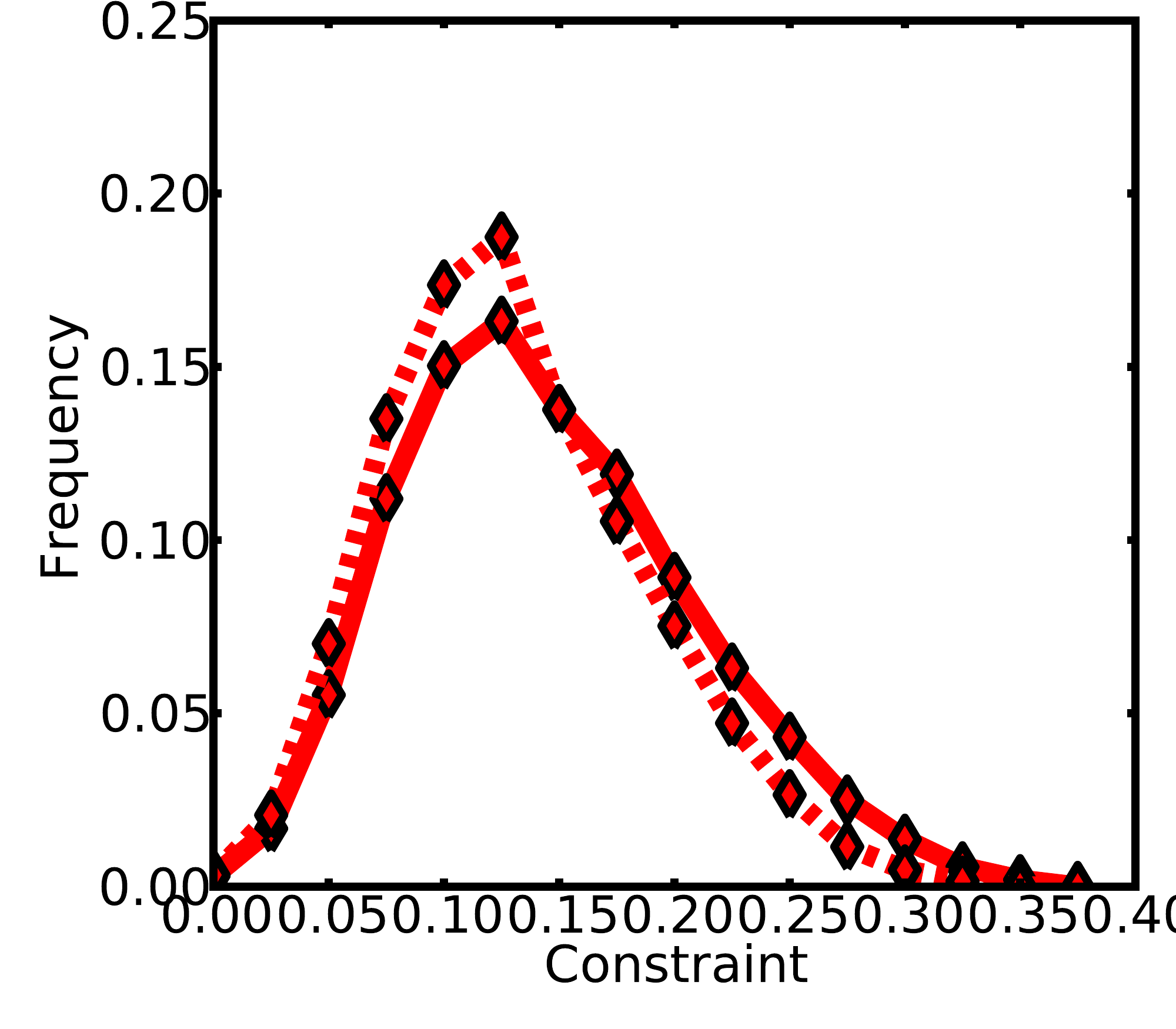} \\
  \hline \hline
  \multicolumn{5}{c}{\textbf{Computer Science}}\\
  \hline \hline \\
  \includegraphics[width=0.37\columnwidth]{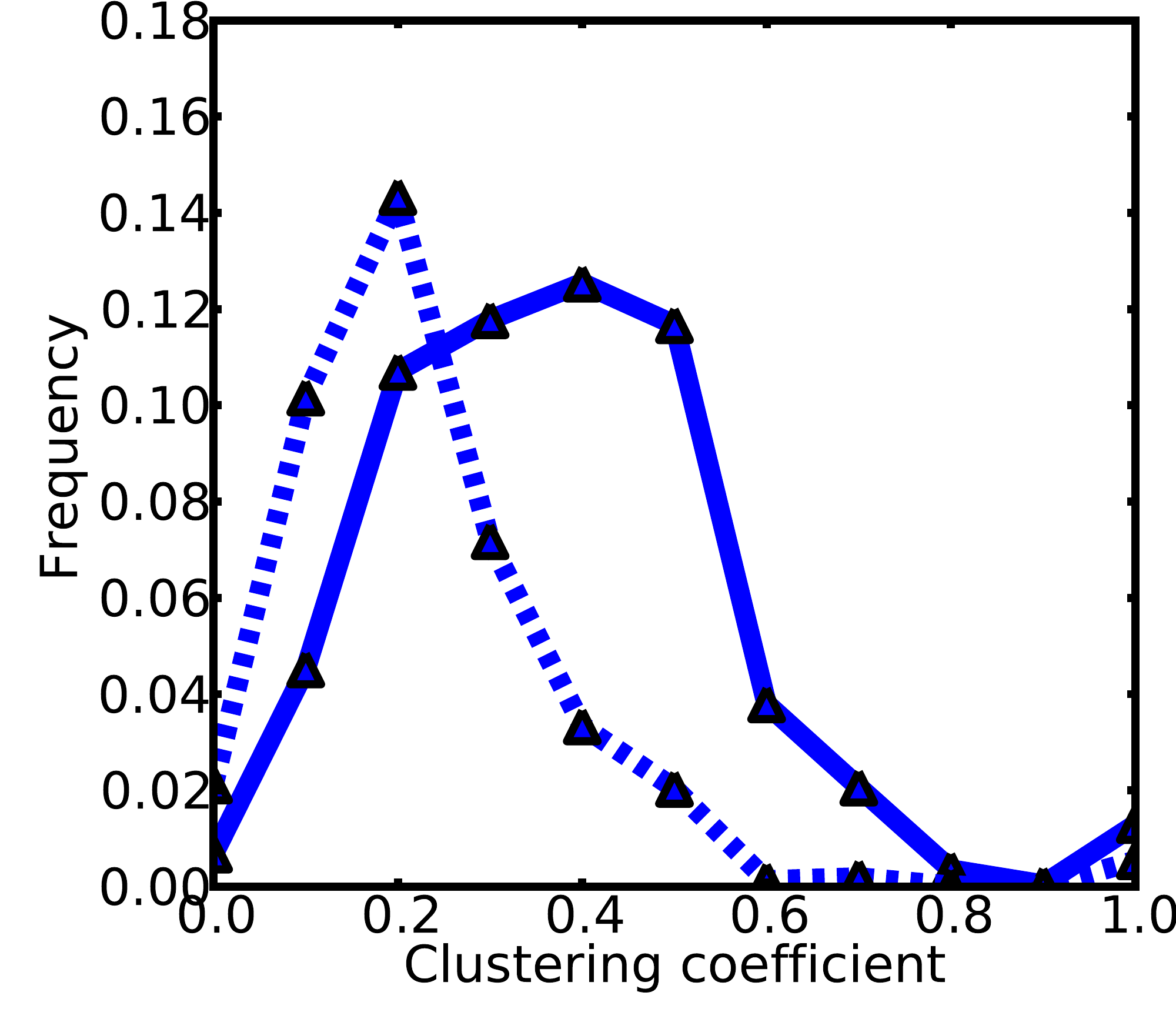} &
  \includegraphics[width=0.37\columnwidth]{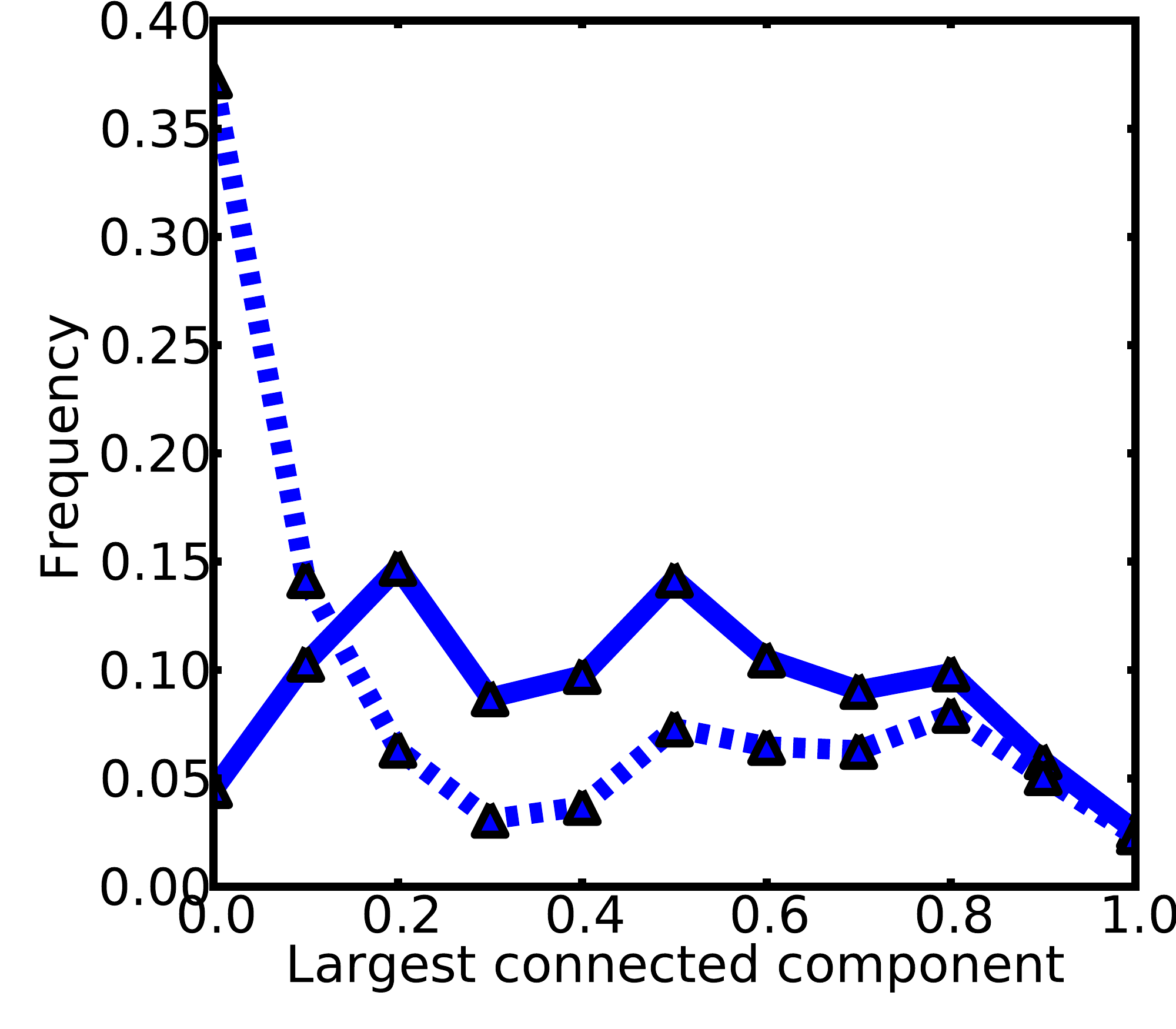} &
  \includegraphics[width=0.37\columnwidth]{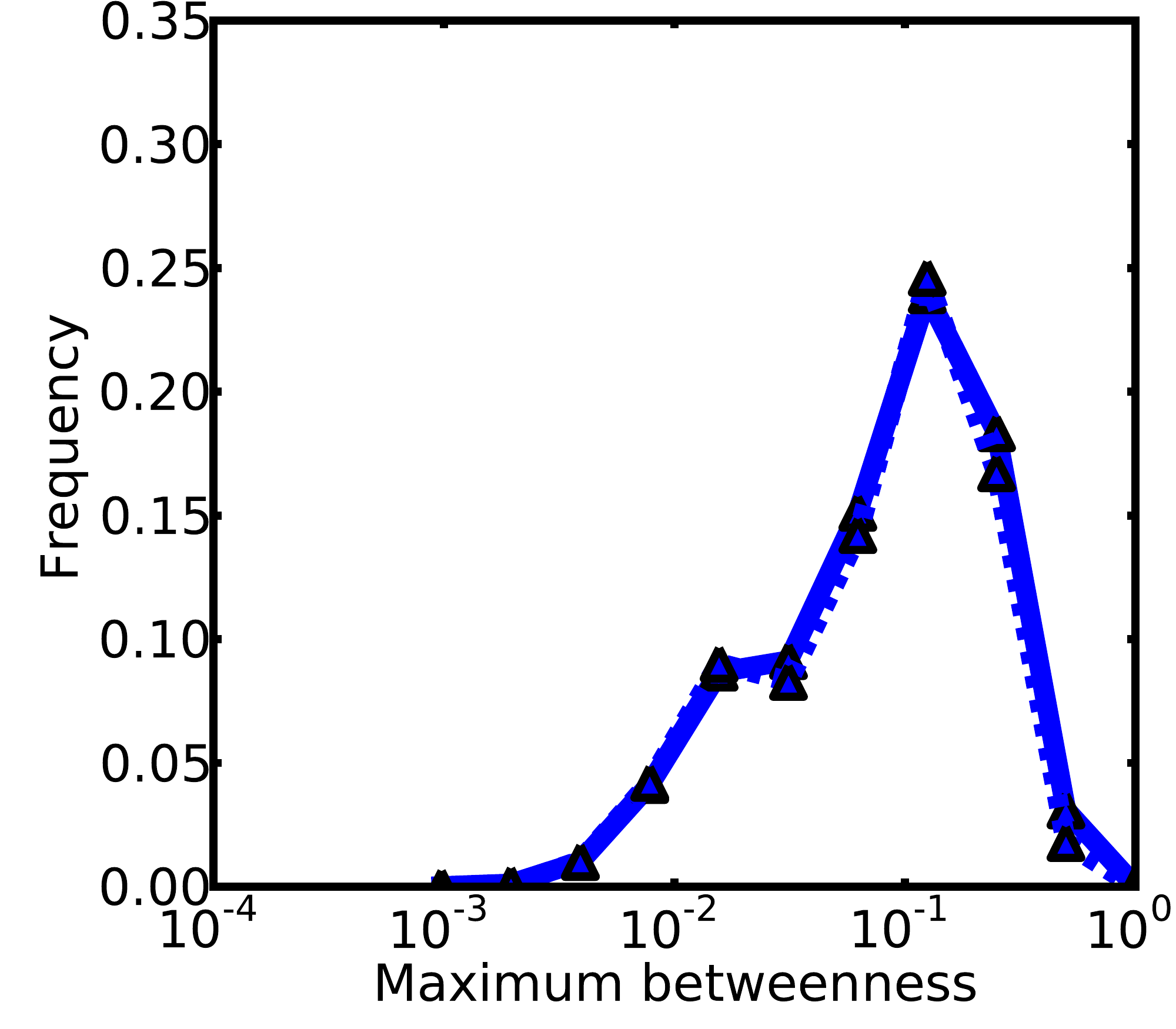}  &
  \includegraphics[width=0.37\columnwidth]{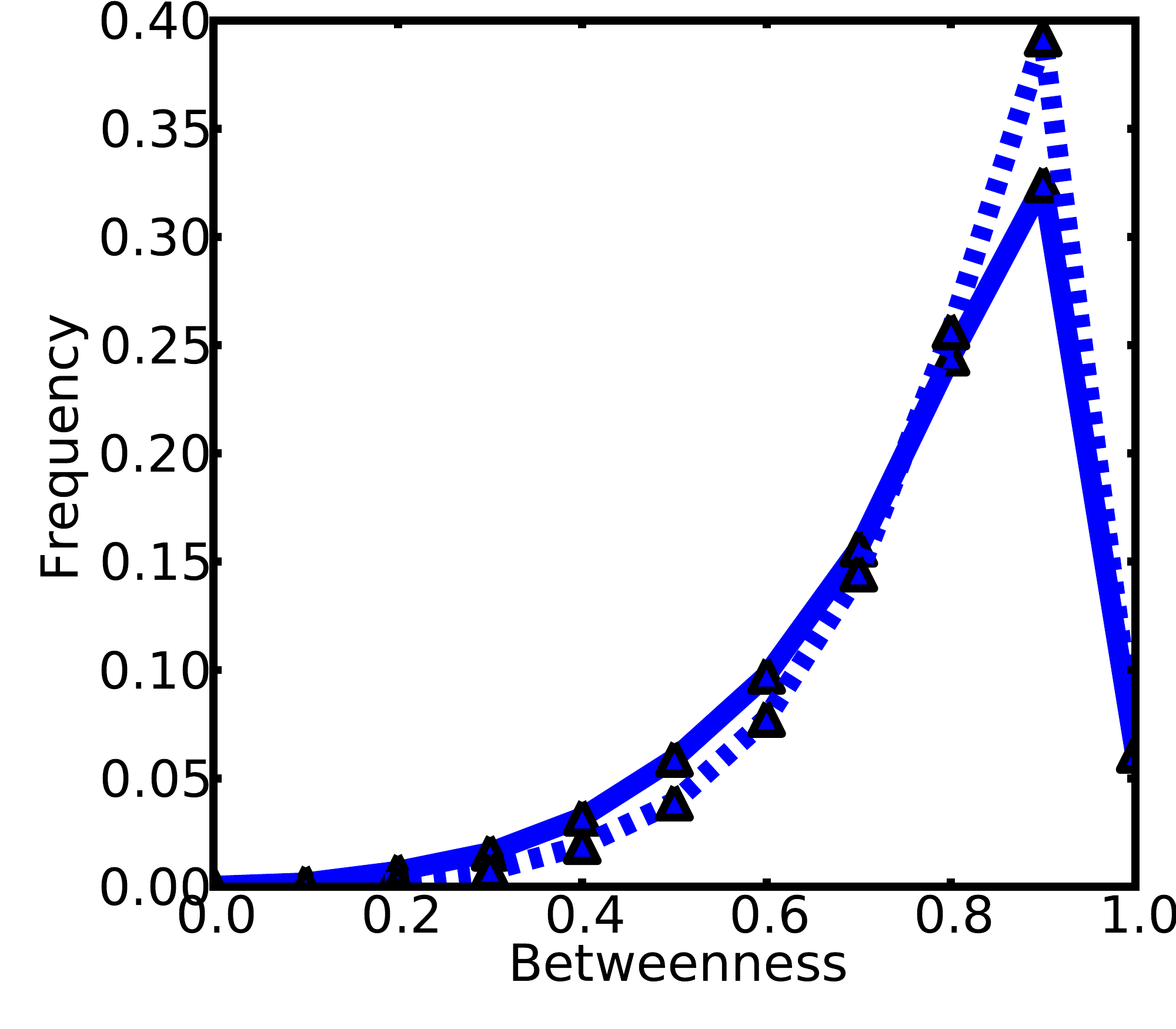}  &
  \includegraphics[width=0.37\columnwidth]{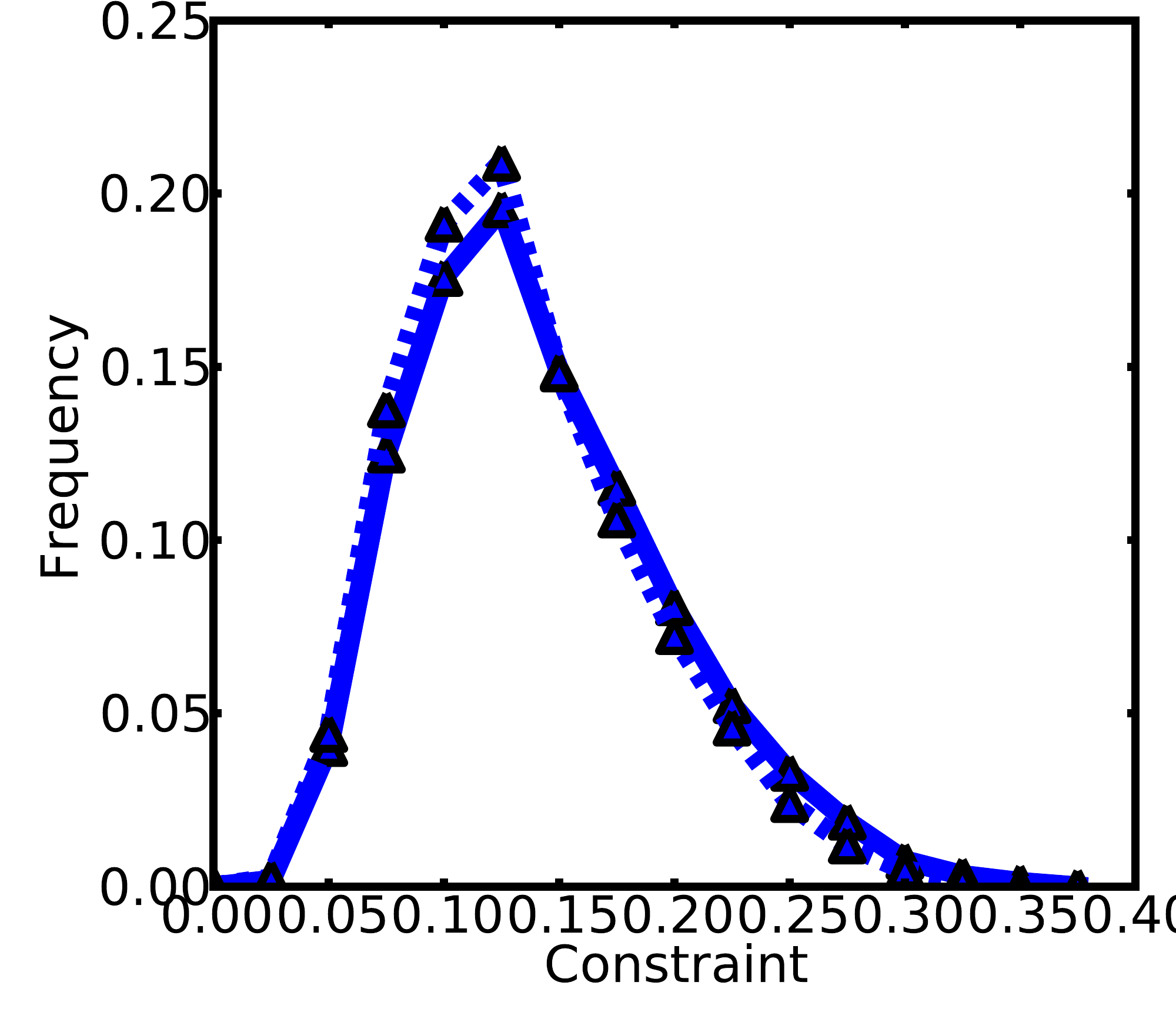}  \\
  M2 & M3 & M4 & M5 & M6 \\
  \end{tabular*}
  \caption{Properties of citation projection graphs in natural science (green), social science (red) and computer science
  (blue) compared with random graphs with same degree sequences. Solid lines represent properties of real citation
  projection networks, while dashed lines represent their randomized counterparts. Note that real citation projection
  networks are much more connected and clustered, which suggests the importance of social capital in a sub-field. $t$-tests show that all the differences are statistically significant.}
  \label{fig:randomgraphs}
%\vspace{-2mm}
\end{figure*}

We begin by comparing properties of citation projection graphs across the three disciplines of science. In particular, we focus on computer science, natural science and social science. We investigate how publications in various areas of science are situated in the citation networks when they got published.

To gain insights into the above question we adopt the following methodology: for every category we take all the papers and for each of them we create its citation projection graphs $G_p$ and $G_{p0}$ and compute the six network metrics M1-M6 defined in Section~\ref{sec:cpg}. These metrics characterize the shape of the citation projection network. For each metric we create a normalized histogram and by comparing these histograms across the disciplines we gain insight into disciplinary norms and strategic placement of papers in the area.

Figure~\ref{fig:areas} gives normalized histograms of the six metrics that characterize the shape of citation projection networks. Each plot has three curves, one for each area of science. The rest of the paper will use consistent color coding. Computer science is always denoted by blue lines with triangles, natural science is denoted by green lines with squares and social science is by red lines with diamonds.

In general, various network metrics seem quite consistent across the areas of science. For example, network density (Figure~\ref{fig:areas}(a)), maximum betweenness (Figure~\ref{fig:areas}(d)) and also betweenness of $v_0$ (Figure~\ref{fig:areas}(e)) are similar between the scientific disciplines. However, the metrics also reveal some important differences in the structure of citation projection networks. For example, citation networks in natural science are generally more focused. This is revealed by higher clustering coefficient (Figure~\ref{fig:areas}(b)), bigger largest connected component (Figure~\ref{fig:areas}(c)) and larger network constraint (Figure~\ref{fig:areas}(f)). This means that in general natural science papers tend to create citation networks that are somewhat focused on a narrow subdiscipline and thus the network contains a relatively large component with many local closed triangles. This is further supported by the larger network constraint of $v_0$.

On the other hand, properties of social science and computer science citation networks are much more similar. While clustering of computer science publications seems to be the smallest, the network constraint of computer science citation networks is also smaller than that of natural science. This indicates that computer science, and to some extent social science papers in general, have more diverse citation projection networks, where they cite a wide range of papers that then do not refer to one another.

These results are interesting, as one would expect computer science to more closely resemble natural science than social science. Also the publication norms and standards intuitively seem more aligned with natural than social science. However, these results suggest that it is the narrow focus of natural science citation projection networks that makes them different from computer and social sciences. This seems to suggest that natural science is very specialized and separated into many small sub-fields where most citations are among the papers between a particular subfield.

These observed differences between disciplines of science raise the interesting question of how ``random'' the citation projection networks are. We would like to know how much of the variability in the network metrics is due to diversity between the subfields of a discipline and how much is simply due to randomness. This is exactly what we investigate next.

\section{Citation Projection Graphs vs. Random Graphs} \label{sec:randomgraphs}

As motivated above we now compare real citation projection graphs with their ``random'' counterparts. This approach will give us insights into how real citation projection networks differ from random ones. Are citation projection graphs more random or more clustered? Are publications more likely to be bridges and structural holes or more likely to be situated within communities? In order to answer these questions we adopt the following approach.

For every citation projection graph $G_p$, we construct a random graph $G_r$ with the same in- and out-degree sequence as those of $G_p$. This means that every $G_p$ now has a corresponding graph $G_r$ with the same number of nodes and same number of edges, and each node also has the same degree (the same number of citations) as in the original $G_p$. This means that the density and the degree sequence of $G_r$ are exactly the same as those of $G_p$. As in the previous section we characterize each $G_p$ and $G_r$ with the 5 metrics (we skip Density (M1) as it remains unchanged between $G_p$ and $G_r$). We then plot the normalized histograms of values and plot the corresponding curves in Figure~\ref{fig:randomgraphs}. We show the distributions of clustering coefficient, largest connected component, maximum betweenness, betweenness of $v_0$, and constraint of $v_0$ in citation projection graphs and their corresponding random graphs.

Comparing plots in Figure~\ref{fig:randomgraphs}, we see that real citation projection networks $G_p$ differ from their corresponding randomized counterparts $G_r$ in a very consistent manner. Publications in all three fields tend to cite previous work of better connectivity (i.e., larger connected component) and in more clustered communities (i.e., higher clustering coefficient). Figure~\ref{fig:randomgraphs} also shows that the random graphs tend to have slightly larger maximum betweenness. Finally, by comparing the distribution of betweenness and network constraint of $v_0$, we see that publications in all these three areas tend to have slightly lower betweenness and higher network constraint than what is expected to be in random graphs. All these facts together suggest that scientific works are usually built upon previous published work of closed, dense and cohesive communities. The differences are especially striking in the size of a largest connected component and clustering coefficient, which suggests that in real citation networks even inside large connected components there are small densely connected sets of papers.

These results as a whole align nicely with social science arguments which suggest that forming links within tightly connected communities helps to increase one's social capital and consequently one's performance. As one is able to give rise to a sense of belonging, this enables the creation of a common culture, and enforces social norms~\cite{Reagans2003}. By studying the citation projection graphs of scientific publications, we see that this tendency consistently holds for scholarly works in most areas in science.

\section{Citation projection graphs and publication impact}\label{sec:diffimpact}

\begin{figure*}[t]
  \centering
  \begin{tabular*}{\textwidth}{@{\extracolsep{\fill}}ccc}
    \includegraphics[width=0.667\columnwidth]{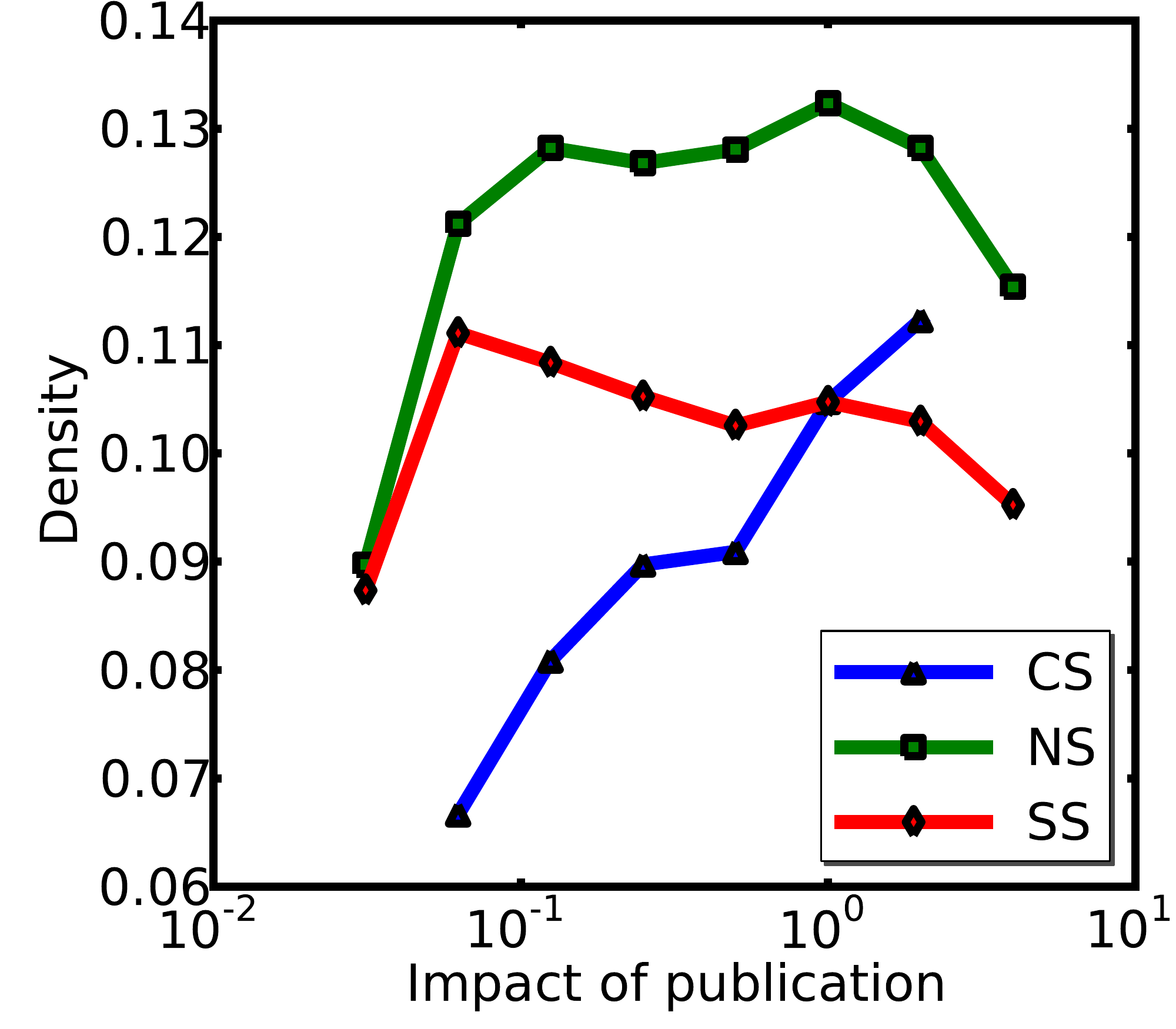} &
    \includegraphics[width=0.667\columnwidth]{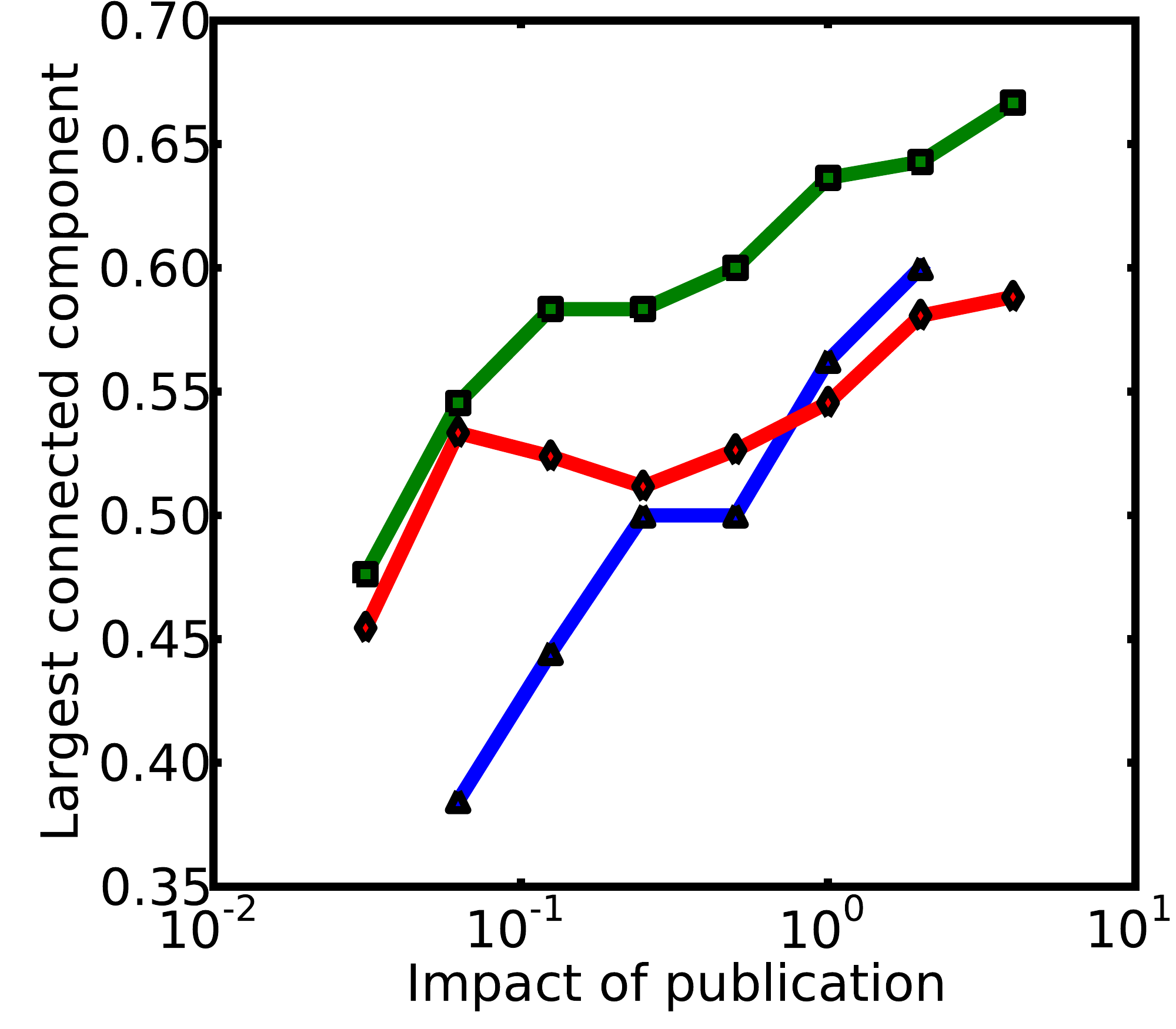} &
    \includegraphics[width=0.667\columnwidth]{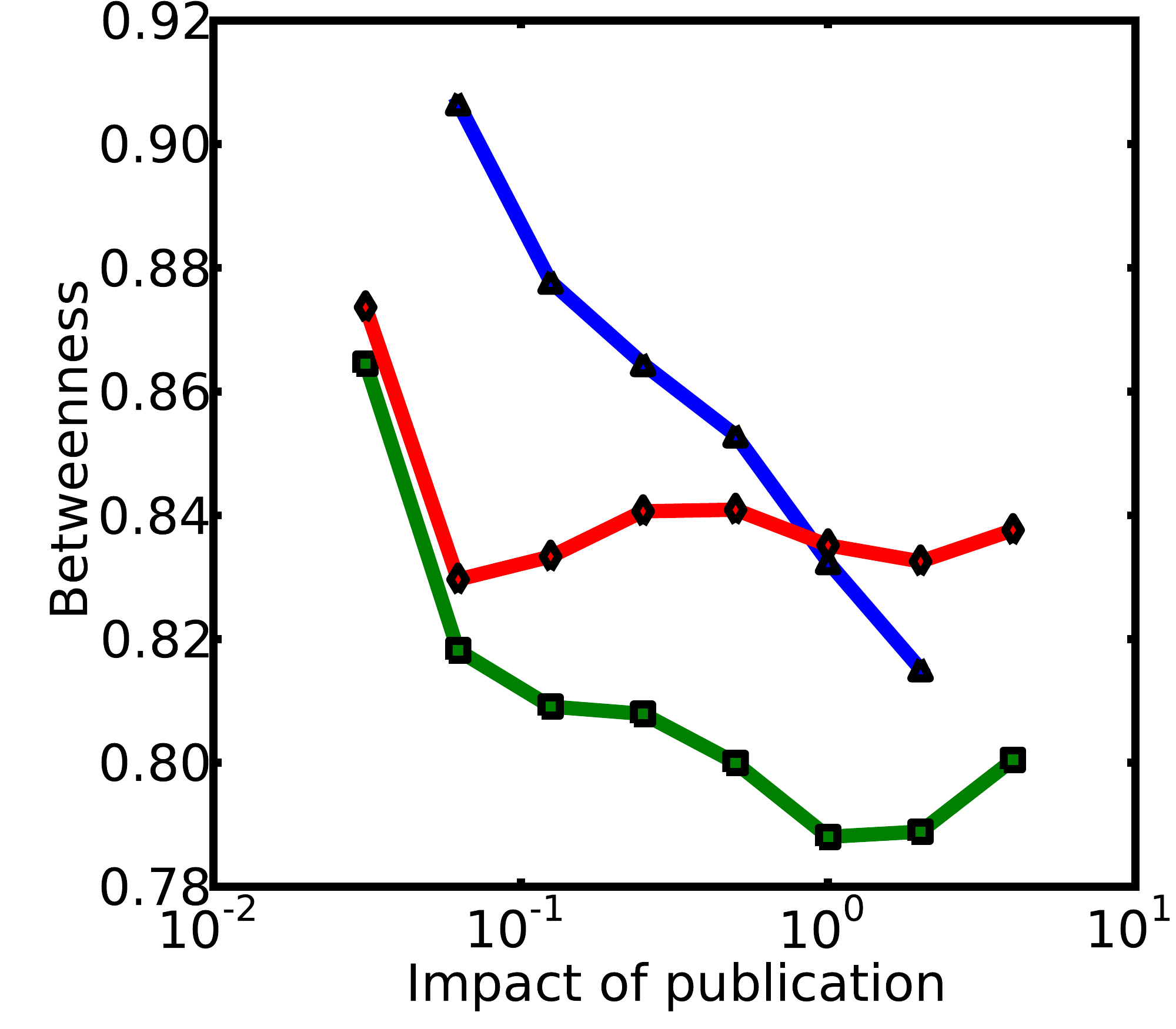}  \\
     (a) M1: Density & (c) M3: Connectivity & (e) M5: Betweenness \\
    \includegraphics[width=0.667\columnwidth]{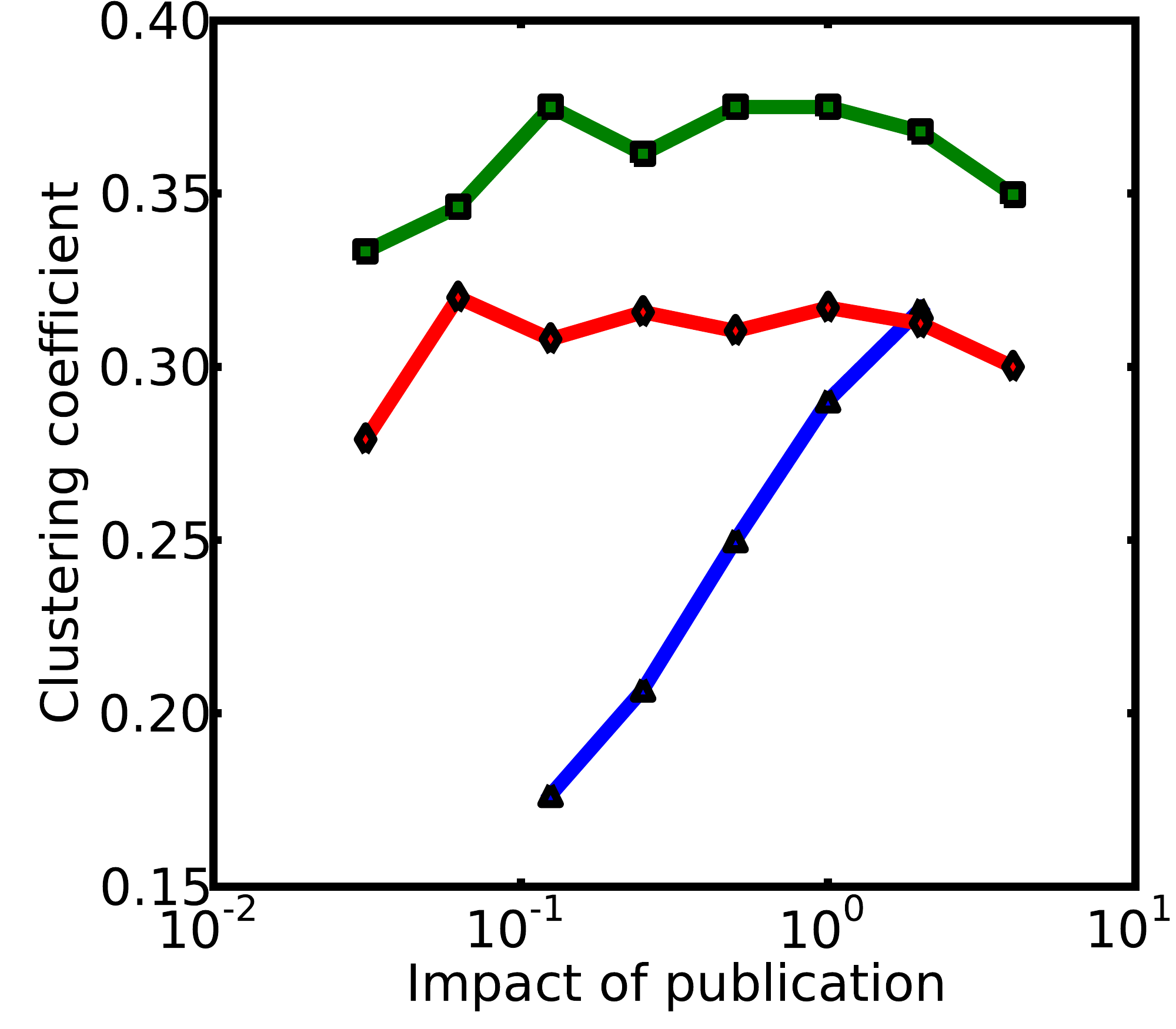} &
    \includegraphics[width=0.667\columnwidth]{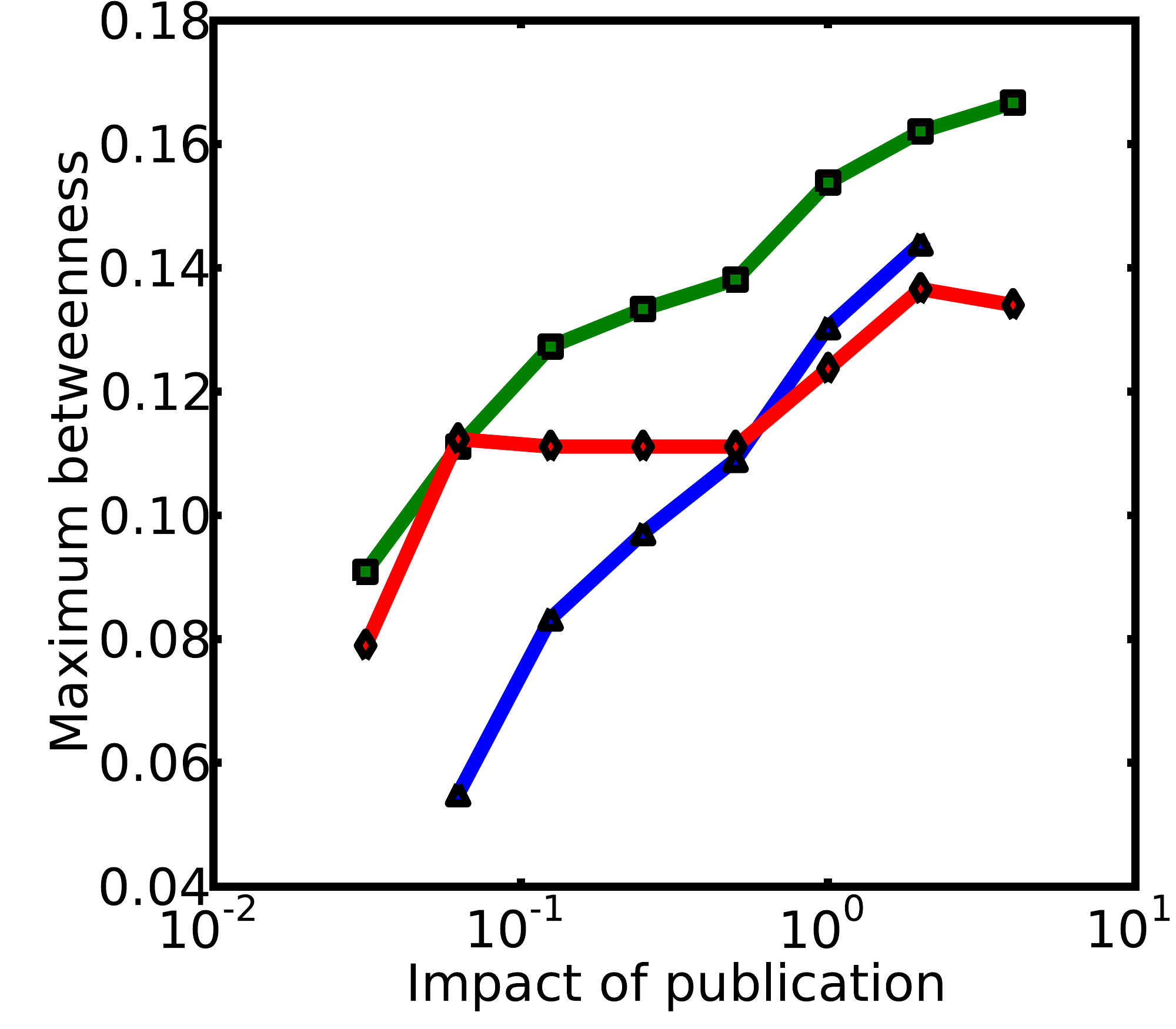} &
    \includegraphics[width=0.667\columnwidth]{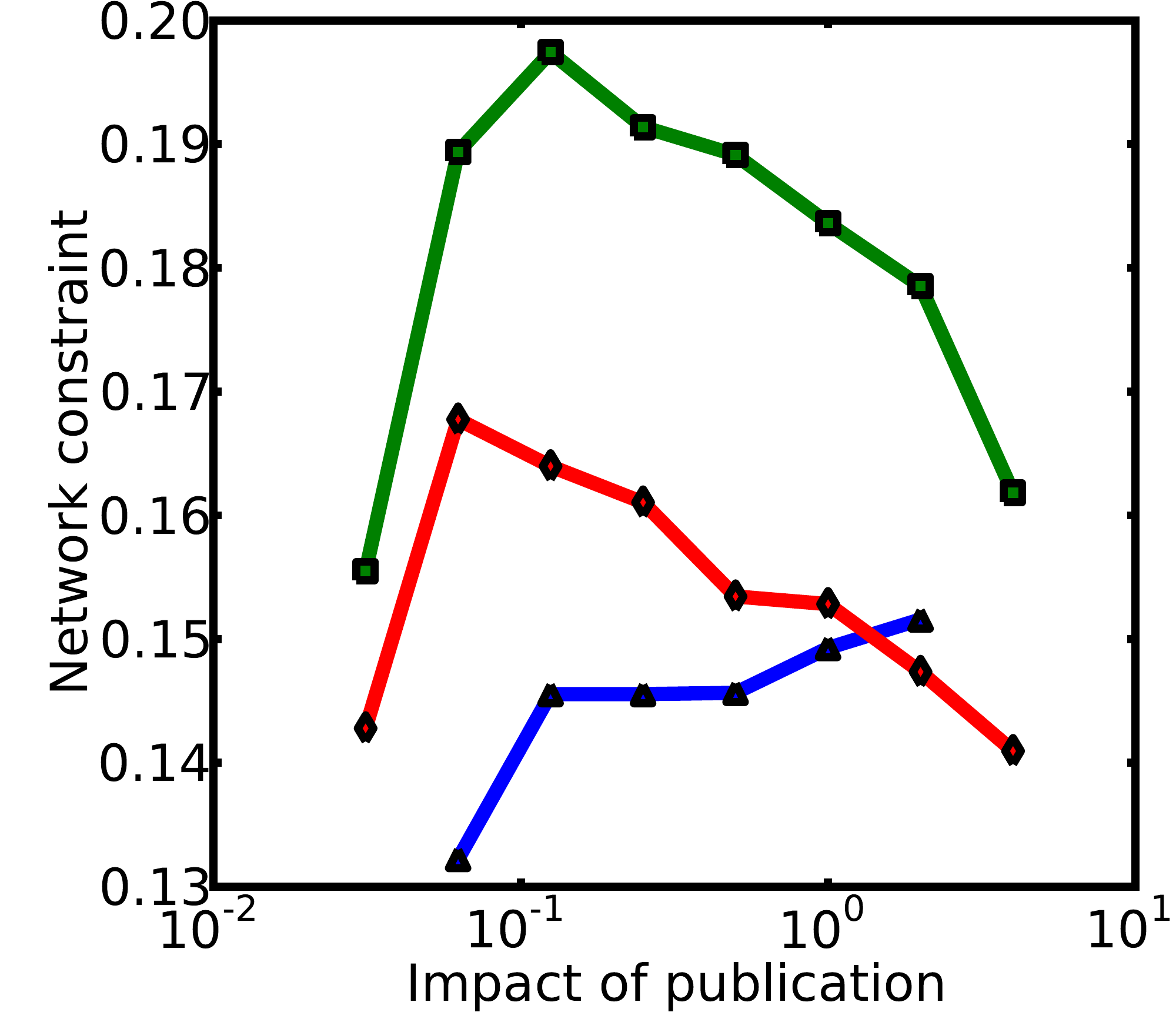}  \\
    (b) M2: Clustering Coefficient & (d) M4: Maximum Betweenness of $v_0$ & (f) M6: Network constraint of $v_0$\\
      \end{tabular*}
  \caption{The median value of a network metric characterizing the citation projection graphs as a function of the
  publication impact. The blue curves with triangles represent publications in computer science, green curves with squares represent natural science,
  and red curves with diamonds are the publications in social science.}
  \label{fig:diffareas}
\vspace{-2mm}
\end{figure*}

Scholarly fields differ, and different disciplinary norms and citation standards give rise to diverse citation projection networks. These norms or styles of citation, in turn, act as contexts within which an article is received and recognized. The question is, do high impact articles conform to disciplinary norms of citation structures or are there certain styles of citation which generate higher impact in spite of these contextual differences? In this section we explore the relation between impact and citation patterns within each discipline.

\subsection{Impact of publications}
There are several different ways of assigning an impact to a paper. The most straightforward way is simply to count the total number of citations a paper received. However, this is not a good measure for two reasons. First, older papers generally have more time to accumulate citations and thus the older the paper is the more ``impact'' it has. The second reason is more subtle. As science changes and evolves, the disciplinary norms, publishing standards and citation patterns also change. Some fields publish very regularly and have long lists of references, while other fields publish rarely and have shorter reference lists. Moreover the scientific output also varies from year to year. We would like to have a measure of impact that will account for all these factors and allow us to objectively compare different scientific disciplines.

We define the {\em impact}~\cite{valderas2007} of a publication $p$ as the number of citations it receives normalized by the average number of citations of all other publications published in the same year and same area as $p$:
%\vspace{-3mm}
$$
I_p = \frac{c_p}{\frac{1}{|P|}\sum_{j \in P}c_j}
$$
where $c_p$ is the number of citations publication $p$ receives, $P$ is the set of publications that appeared in the same year and same area as $p$. This measure allows us to make a fair comparison between articles that may not yet have finished accumulating citations due to their recency and to account for differences in size and publication cycle for different disciplines~\cite{stringer2008ejr}. In Section~\ref{sec:datadescription}, we stated that all the citations we use in this study are within the datasets. Although this could skew the observed raw citation counts toward disciplines that are better represented within each dataset, the normalization by discipline helps to mitigate such biases.

\subsection{Graph patterns vs. impact}\label{sec:patternimpact}

Having defined a measure of publication impact we are now able to correlate the impact of the publication with the structure of the citation projection network. We are interested in whether publications of different impacts also exhibit differences in their citation projection networks.

In Figure~\ref{fig:diffareas} we plot how a particular property of a citation projection network varies with the impact of the publication. Note that the variances of the network property values are very large and highly skewed. In order to gain a fair comparison, we plot the median of the network property value for each value of impact.

Several interesting observations can be made here. In Figure~\ref{fig:diffareas}(a), we see that publications in natural science have the most dense citation projection graphs, which supports the findings of previous sections. Moreover, we also notice the difference between natural and social science networks and computer science networks. In computer science, high impact publications are characterized by relatively dense citation projection networks, while low impact publications cite very idiosyncratically and thus have less dense citation networks. For publications in natural and social sciences, Figure~\ref{fig:diffareas}(a) suggests that medium impact publications have relatively dense networks, while low and high impact publications have sparser citation networks.

Figure~\ref{fig:diffareas}(b) indicates that publications with different impact in natural science and social science have projection citation graphs of about the same clustering level, while for publications in computer science the clustering level increases as the impact increases. This further suggests that highly cited works in computer science tend to have focused citation networks with a large, densely connected component and lots of local clustering.

In terms of the size of the largest connected component and the maximum betweenness in the network (Figures~\ref{fig:diffareas}(c) and (d)) we see similar trends across all three disciplines. High impact publications tend to have more nodes in the largest connected component and better connectivity in terms of the maximum betweenness. This suggests that high impact publications in all three areas are more likely to cite bridges across reference communities. These trends are consistent across the natural, social and computer sciences and are especially pronounced for natural and computer science publications while they are weaker in social science.

Figure~\ref{fig:diffareas}(e) shows a very interesting trend. Betweenness of $v_0$ basically measures what fraction of shortest paths between cited papers pass through $v_0$. If a paper cites idiosyncratically then $v_0$ will have very high betweenness as the only way to get from one cited paper to another is through $v_0$. We observe that this is the case. Low impact publications cite very idiosyncratically and thus they have very high betweenness in their citation projection network. In social science the trend then stabilizes, which suggests that medium and high impact publications have about the same betweenness. Consistent with our previous findings high impact publications in computer science have very low betweenness in their citation projection graph. This means that shortest paths between cited publications pass directly through the projection graph rather than to go through $v_0$.

Last, the shape of the network constraint curve (Figure~\ref{fig:diffareas}(f)) resembles the patterns of the network density measure (Figure~\ref{fig:diffareas}(a)). Publications in natural science have the highest network constraint, and such constraint is relatively low for both low and high impact papers. Publications in social science share a similar trend, although the constraint is much lower. However, publications in computer science have even lower network constraint. Again note that except in computer science, high and low impact publications have similar median value of network constraint, while in computer science the higher impact publications are more embedded into a small, densely connected community.

We see that the left-most points in about half of the curves in Figure~\ref{fig:diffareas} ((a), (e), (f)) behave somewhat differently from the remaining points. This is because the initial points include lots of  zero-impact publications -- publications that are never cited after being published. This is consistent with the observation in \cite{Shi2009}, which shows that zero-impact publications have properties distinct from other publications. Usually 20\% - 30\% of all publications have no subsequent citations in the datasets, so the properties of these publications affect observations in the entire citation network.

All in all we note that publications in computer science, natural science and social science are likely to have dense, well-connected citation projection graphs. We also see that the impact of publications in computer science is linearly correlated with most of the properties, such as the graph density,  clustering coefficient, connectivity and betweenness; while publications in other areas do not have this trend.

These experiments raise a very interesting question. It seems that medium impact publications cite in a focused manner, with most citations going to within-community papers (e.g., as in Figure~\ref{fig:examples}(b)), while high and low impact publications seem to create networks of lower density and lower network constraint, which suggests that their projection networks look more like the examples in Figure~\ref{fig:examples}(a) and (c). Next we focus on exactly this question and further examine the differences in the citation projection networks between high and low impact publications.

\subsection{Citation patterns of high and low impact publications}

\begin{table*}[t]
  \centering
   \caption{Average citation projection network statistics for the high, medium and low impact publications for the
    three areas of science: computer science (CS), natural science (NS) and social science (SS).
    Columns 3-5 show mean values of the properties, and columns 6-8 give the $p$-values of the statistical
    significance in the means of the two distributions as calculated by $t$-tests.
    The highest values of Density, Clustering Coefficient, Connectivity, Maximum Betweenness, and
    Network Constraint are highlighted, and the lowest values of Betweenness are highlighted. Note the consistency
    between natural and social science, whereas computer science seems to be a small outlier.}
    \vspace{2mm}
  %\resizebox{!}{!} {
  \begin{tabular}{| x{2cm} | x{0.8cm} | x{1.3cm} | x{1.3cm} | x{1.3cm} | x{1.9cm} | x{1.9cm} | x{1.9cm} | }
  \hline
  \rowcolor[gray]{0.8} Feature & Area & High & Mid & Low & High vs. Mid & Mid vs. Low & High vs. Low \tabularnewline
    \hline \hline
    Density & CS & \textbf{0.132} & 0.117 & 0.094 & 9.44e-08 & $<$2.2e-16 & $<$2.2e-16 \tabularnewline
    (M1) & NS & 0.139 & \textbf{0.150} & 0.114 & 3.52e-15 & $<$2.2e-16 & $<$2.2e-16 \tabularnewline
    & SS &0.116 & \textbf{0.122}  & 0.102 & 2.07e-05 & $< $2.2e-16 & $< $2.2e-16 \tabularnewline
    \hline
    Clustering & CS & \textbf{0.298} & 0.259 & 0.217  & 4.27e-07  &  4.18e-14 & $<$2.2e-16 \tabularnewline
    Coefficient & NS & 0.333 & \textbf{0.344} & 0.305 & 0.002  & $<$2.2e-16 & 8.17e-14 \tabularnewline
    (M2) & SS & 0.284 & \textbf{0.292} & 0.267 & 0.039 &  $<$2.2e-16 & 3.96e-05 \tabularnewline
    \hline
    Connectivity & CS & \textbf{0.579} & 0.520 & 0.432 & 1.82e-13 & $<$2.2e-16 & $<$2.2e-16 \tabularnewline
    (M3) & NS & \textbf{0.628} & 0.597 & 0.498 & 7.58e-16 & $<$2.2e-16 & $<$2.2e-16 \tabularnewline
    & SS & \textbf{0.564} &  0.541 & 0.472  & 2.95e-07 & $<$2.2e-16 & $<$2.2e-16 \tabularnewline
    \hline
    Maximum & CS & \textbf{0.175} & 0.153 & 0.112 & 5.51e-06 & $<$2.2e-16 & $<$2.2e-16 \tabularnewline
    Betweenness & NS & \textbf{0.187} & 0.173 & 0.130 & 1.39e-09 & $<$2.2e-16 & $<$2.2e-16 \tabularnewline
    (M4) & SS & \textbf{0.159} & 0.152 & 0.123 & 0.003 & $<$2.2e-16 & $<$2.2e-16 \tabularnewline
    \hline
    Betweenness & CS & \textbf{0.778} & 0.806 & 0.849 & 4.51e-08 & $<$2.2e-16 & $<$2.2e-16 \tabularnewline
    of $v_0$ & NS & 0.763 & \textbf{0.760} & 0.820 & 0.168 & $<$2.2e-16 & $<$2.2e-16 \tabularnewline
    (M5) & SS & 0.802 & \textbf{0.801} & 0.837 & 0.703 &$<$ 2.2e-16 & $<$ 2.2e-16 \tabularnewline
    \hline
    Constraint & CS & \textbf{0.159} & 0.156 & 0.146 & 0.026 & 6.14e-14 & 7.42e-12 \tabularnewline
    of $v_0$& NS & 0.171 & \textbf{0.192} & 0.165  & $<$2.2e-16 & $<$2.2e-16 & 1.07e-06 \tabularnewline
    (M6) & SS & 0.150 & \textbf{0.162} & 0.151  & $<$2.2e-16 & $<$2.2e-16 & 0.299 \tabularnewline
    \hline
    \end{tabular}
    \label{table:diffimpact}
%\vspace{-3mm}
\end{table*}

In Section~\ref{sec:randomgraphs} we saw that publications tend to situate themselves in more coherent environments. However, in Section~\ref{sec:patternimpact} we also saw that properties of citation projection networks vary considerably between publications of high and low impact. Somewhat surprisingly we saw that high and low impact publications tend to have more similar projection networks, while projection networks of medium impact publications are different. In this part we investigate this observation in greater details by further comparing the properties of citation projection graphs of publications with different impact levels. We use statistical hypothesis testing to gain further insights into the differences in citation patterns between low and high impact publications.

First we divide the publications into three separate groups: high, medium and low impact publications. For each discipline we select the top 10\% of publications with the highest impact and consider them ``high impact publications''; the bottom 25\% of publications with the lowest impact are considered ``low impact publications''; the rest are considered to be ``average.'' Note that most of the low impact publications (more than 90\%) have zero-impact.

Table~\ref{table:diffimpact} gives the average value of each of the six citation projection network metrics of publications in the three scientific disciplines. We also show the $p$-values of the $t$-tests to test whether the mean values of network statistics between high, low and medium impact publications are significantly different.

Overall, and similar to what we have seen in Figure~\ref{fig:diffareas}, we find that computer science behaves somewhat differently from natural and social sciences, where the differences between high, low and medium impact papers are very consistent. Thus we first describe the structure of citation networks of papers in social and natural sciences and defer the discussion of computer science for later.

\begin{figure*}[!t]
  \centering
  \begin{tabular*}{0.8\textwidth}{@{\extracolsep{\fill}}ccc}
    \hline \hline
     \multicolumn{3}{c}{\bf Natural Science} \\
    \hline    \hline \\
  %  \vspace{1mm}\\
    \includegraphics[width=0.5\columnwidth]{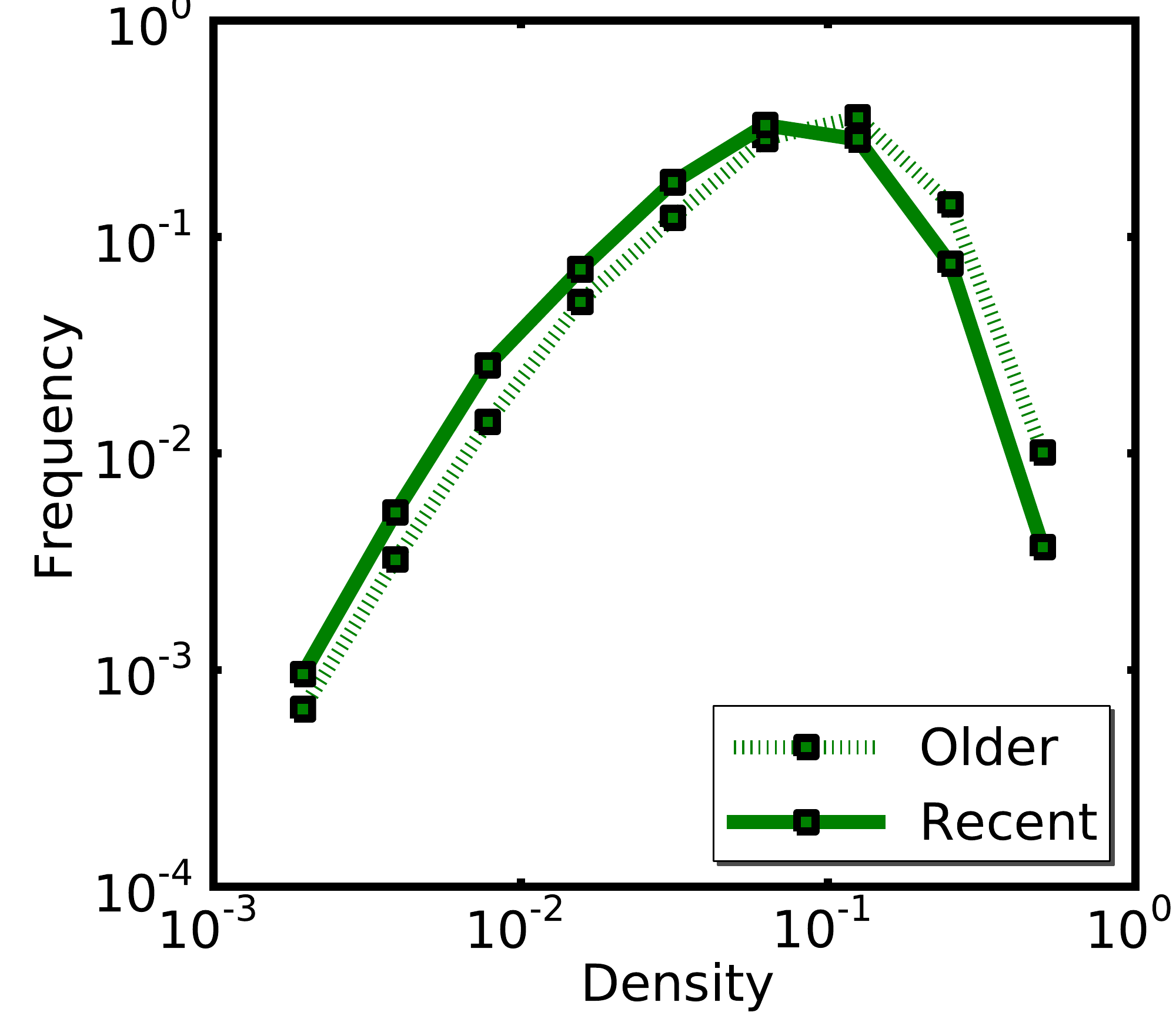} &
    \includegraphics[width=0.5\columnwidth]{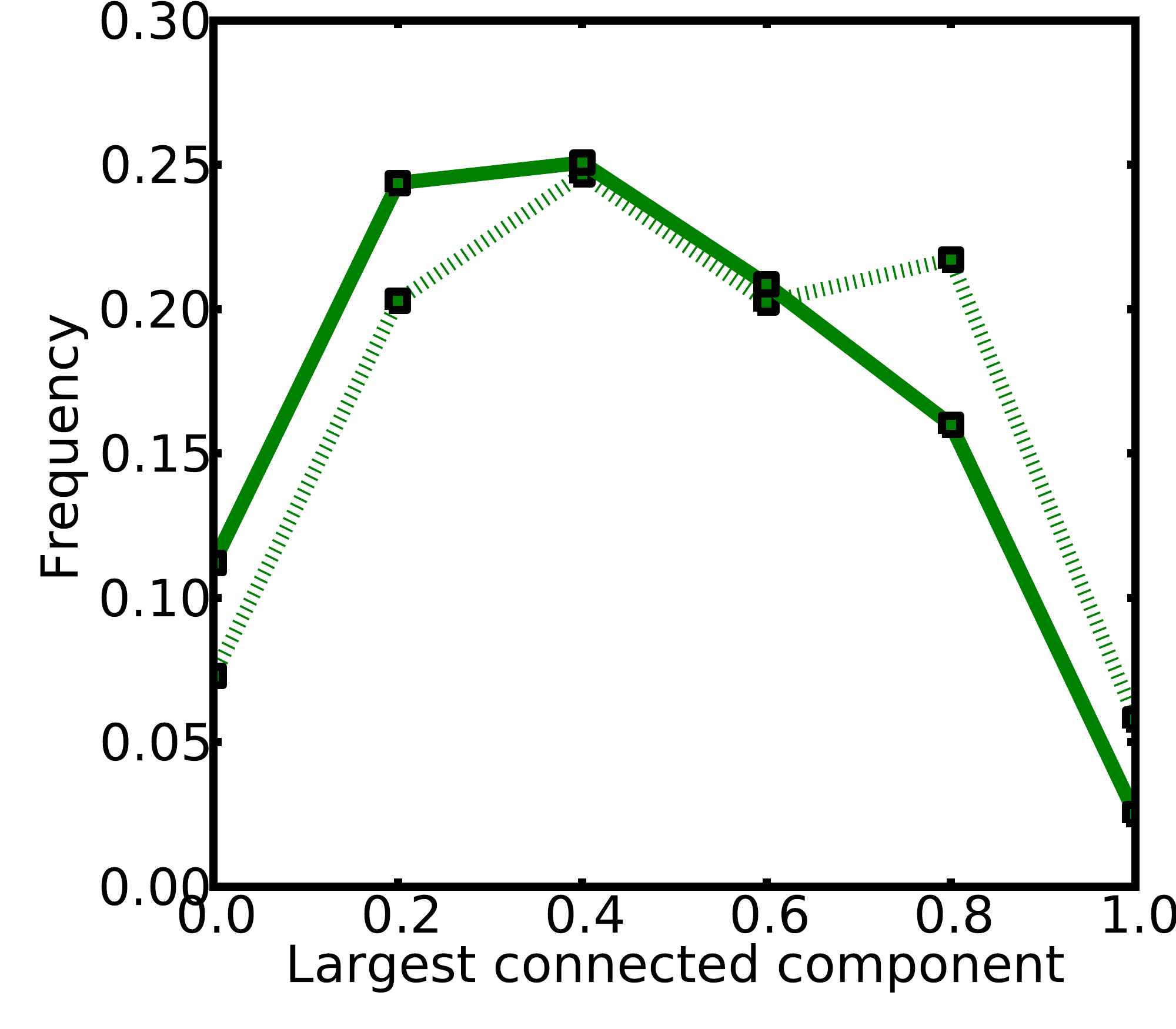} &
    \includegraphics[width=0.5\columnwidth]{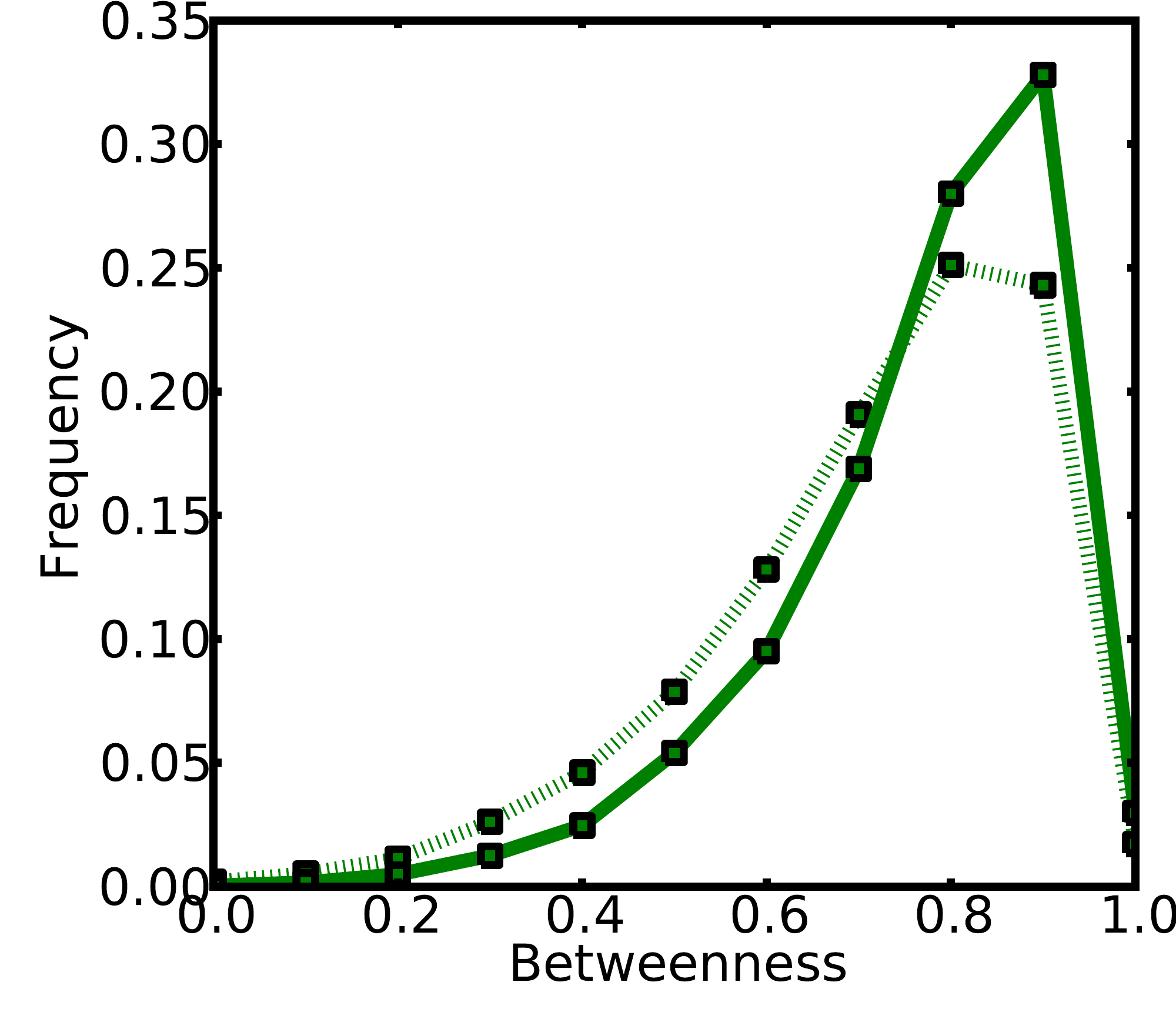} \\
    (a) M1: Density & (c) M3: Connectivity & (e) M5: Betweenness \\
    \includegraphics[width=0.5\columnwidth]{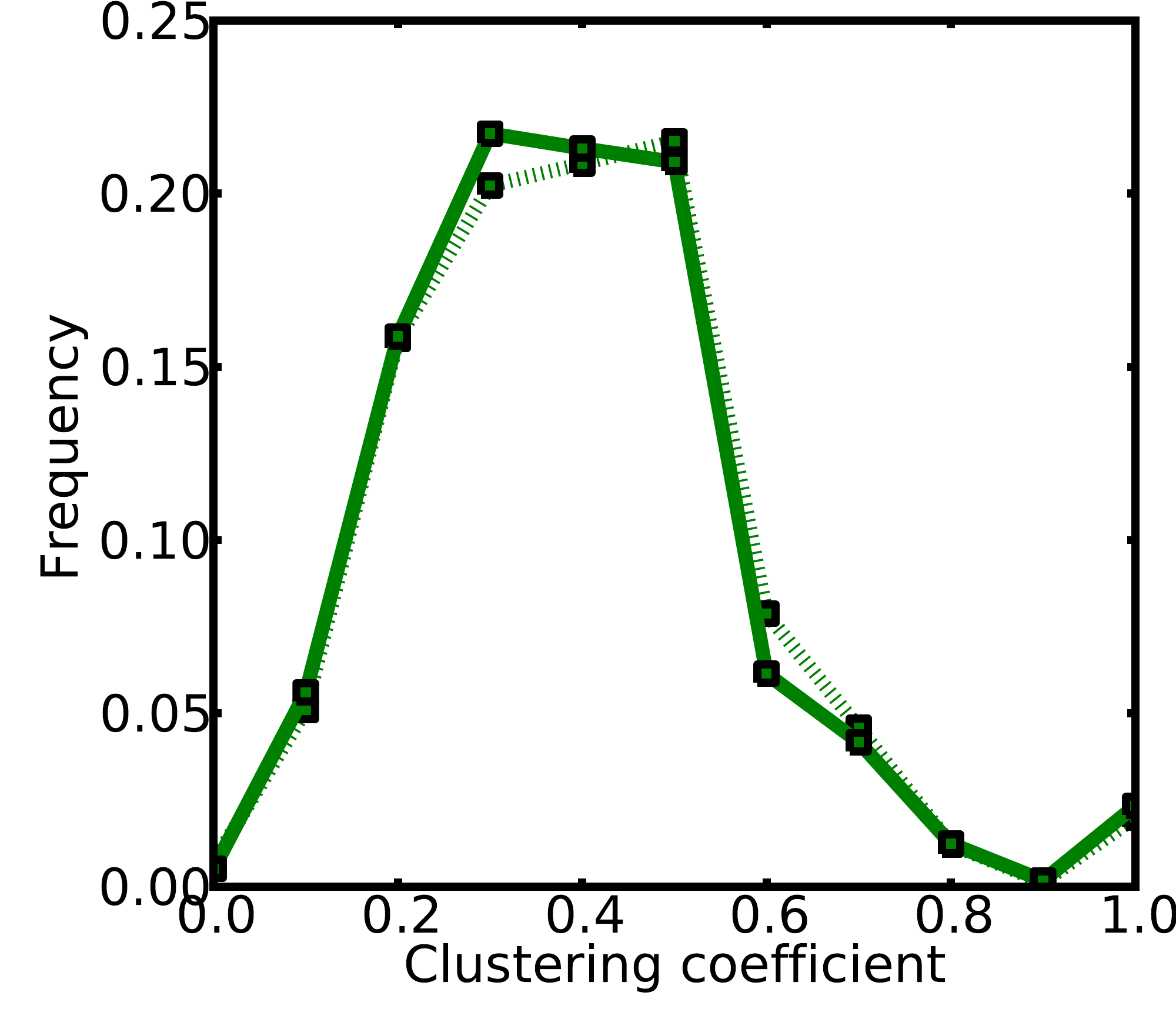} &
    \includegraphics[width=0.5\columnwidth]{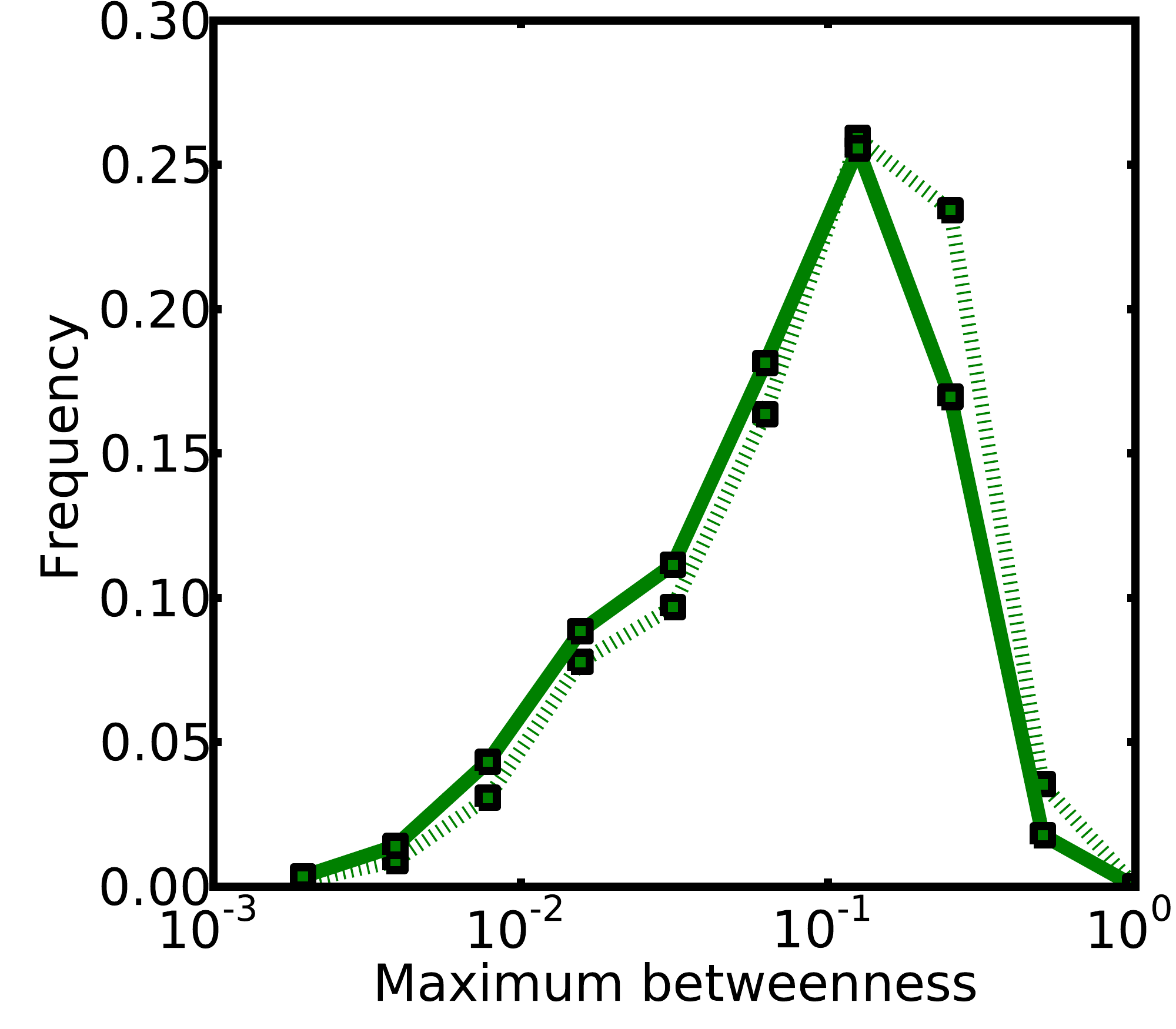} &
    \includegraphics[width=0.5\columnwidth]{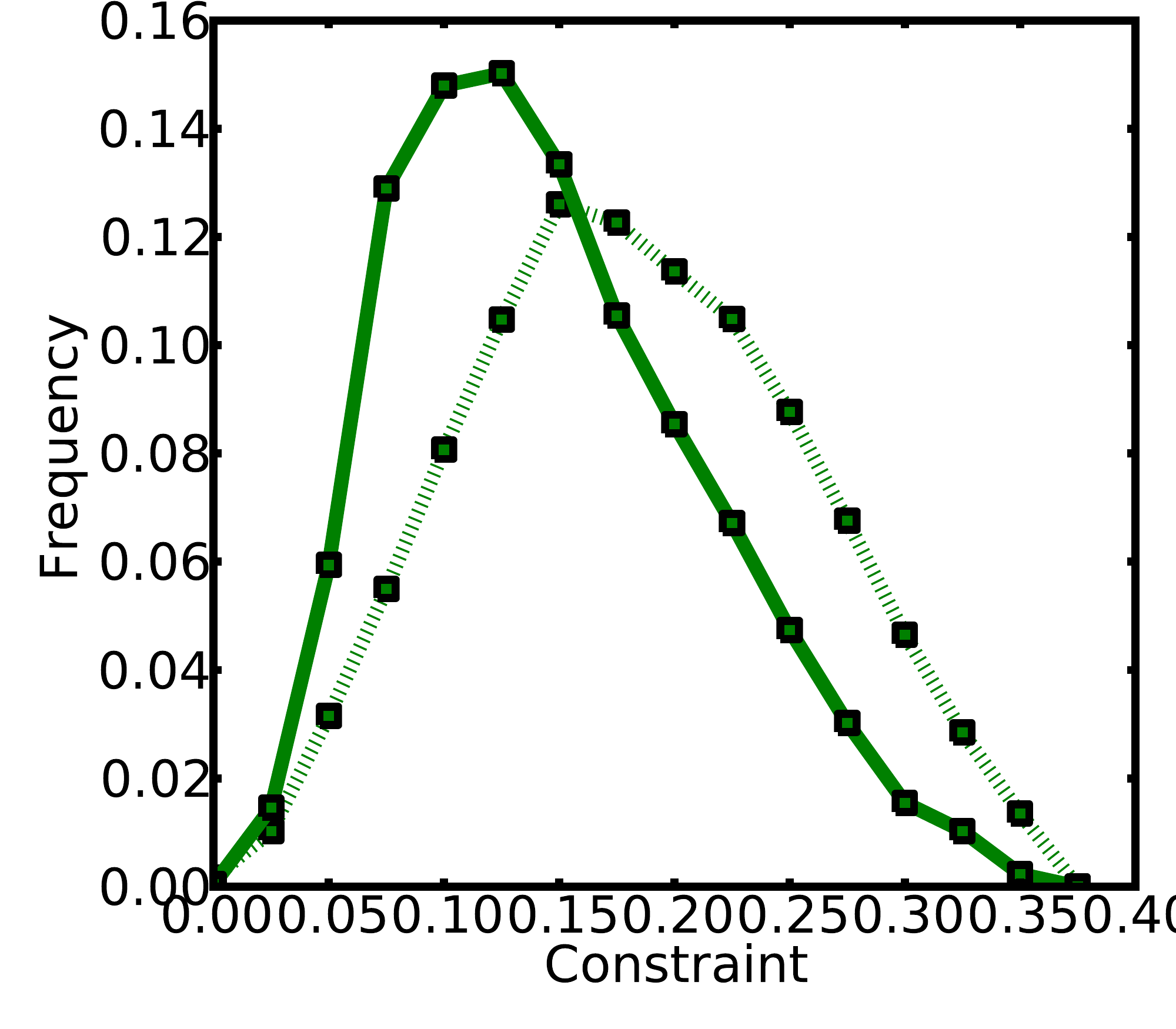}  \\
     (b) M2: Clustering Coefficient & (d) M4: Maximum Betweenness of $v_0$ & (f) M6: Network constraint of $v_0$\\
     \vspace{1mm}\\
    \hline    \hline
    \multicolumn{3}{c}{\bf Computer Science}\\
    \hline    \hline \\
    \includegraphics[width=0.5\columnwidth]{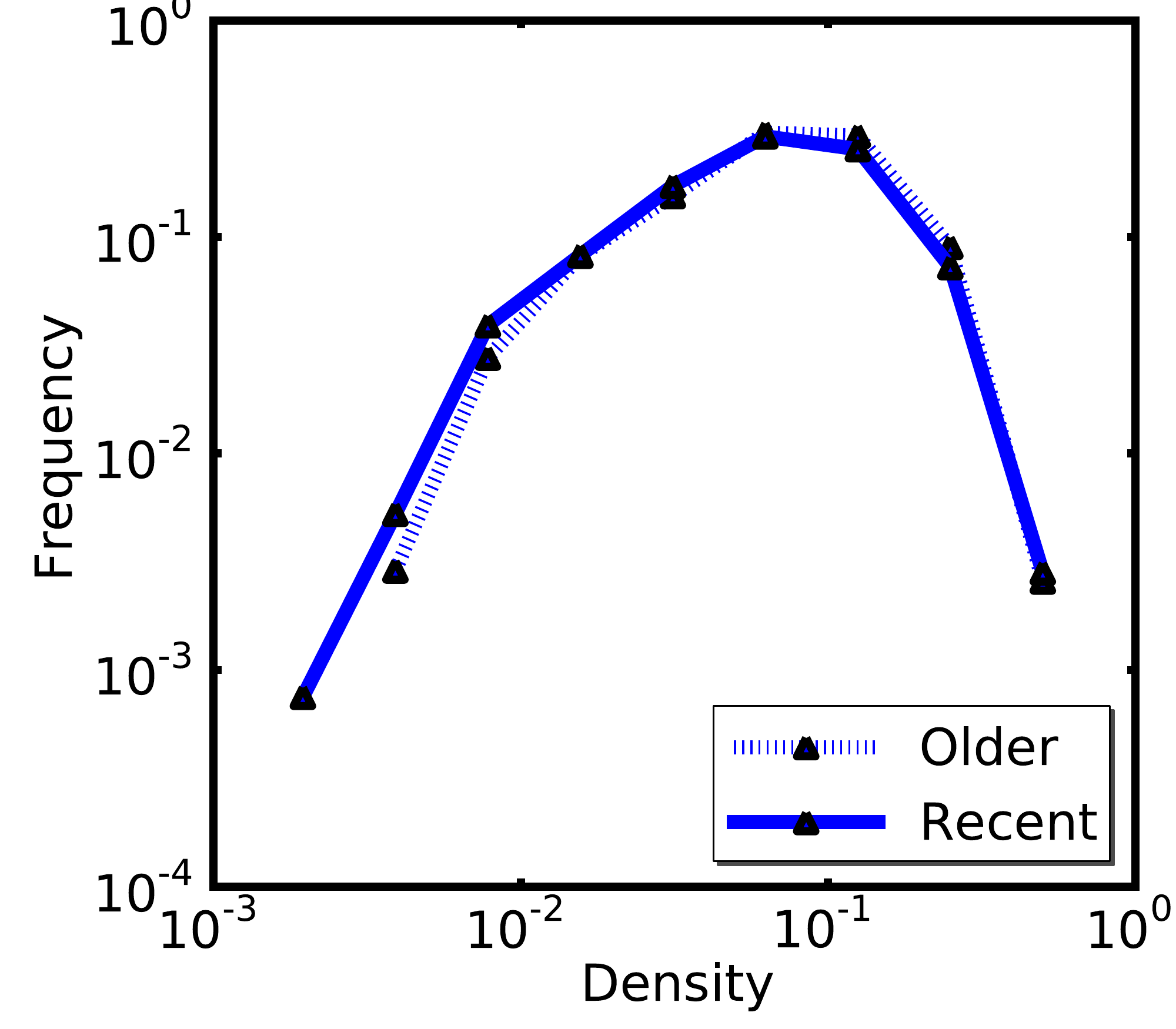} &
    \includegraphics[width=0.5\columnwidth]{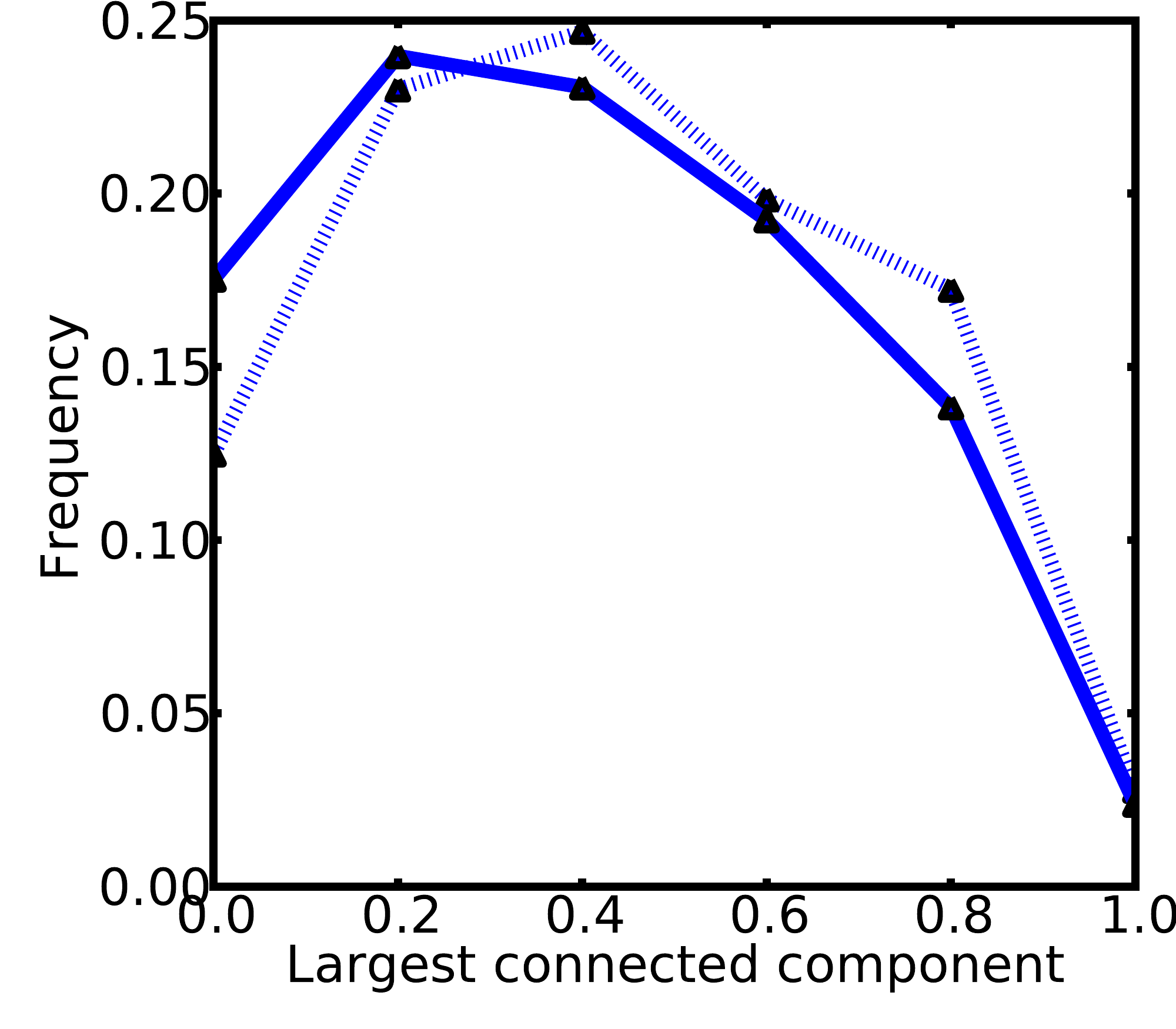} &
    \includegraphics[width=0.5\columnwidth]{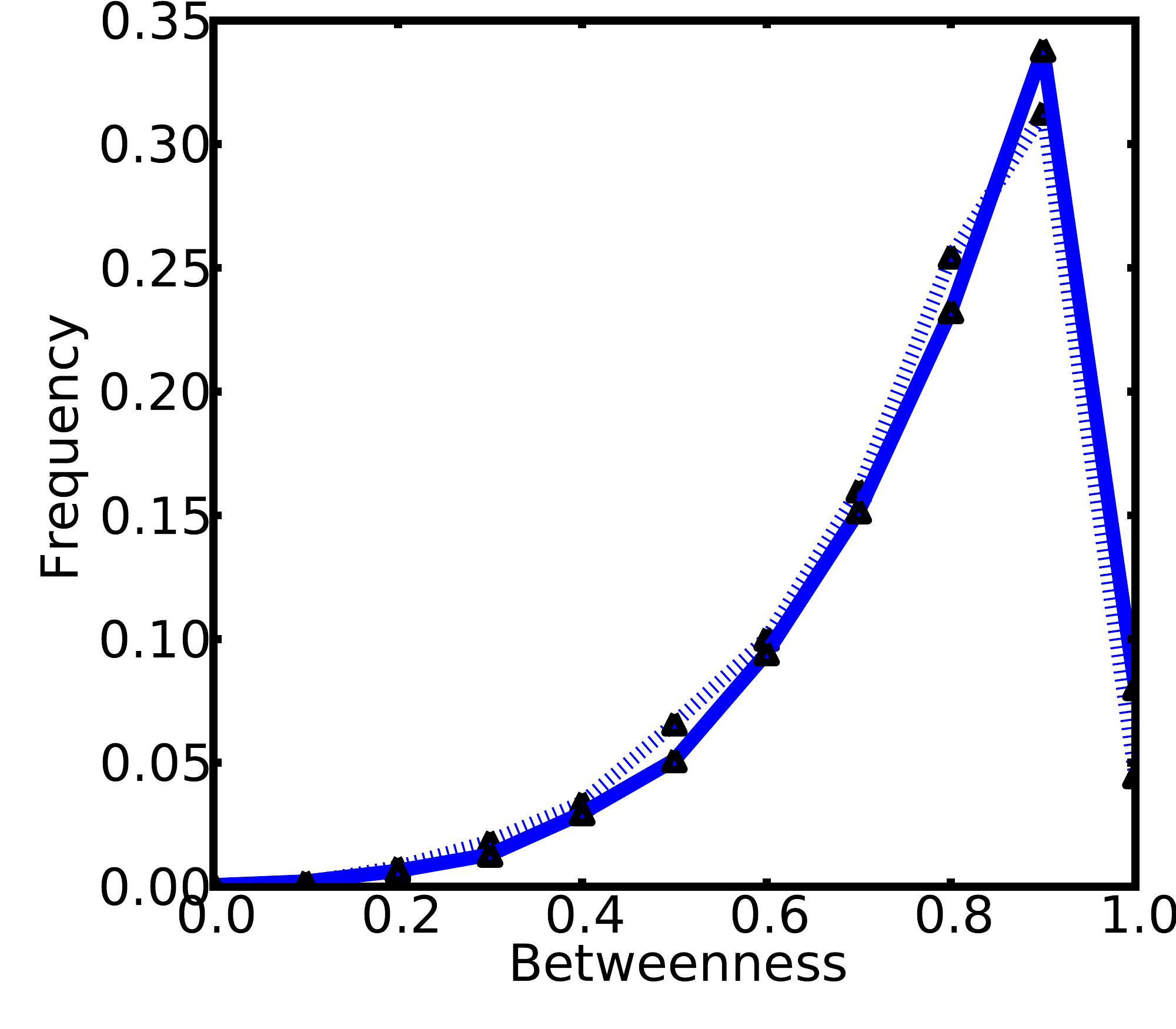} \\
   (a) M1: Density & (c) M3: Connectivity & (e) M5: Betweenness \\
    \includegraphics[width=0.5\columnwidth]{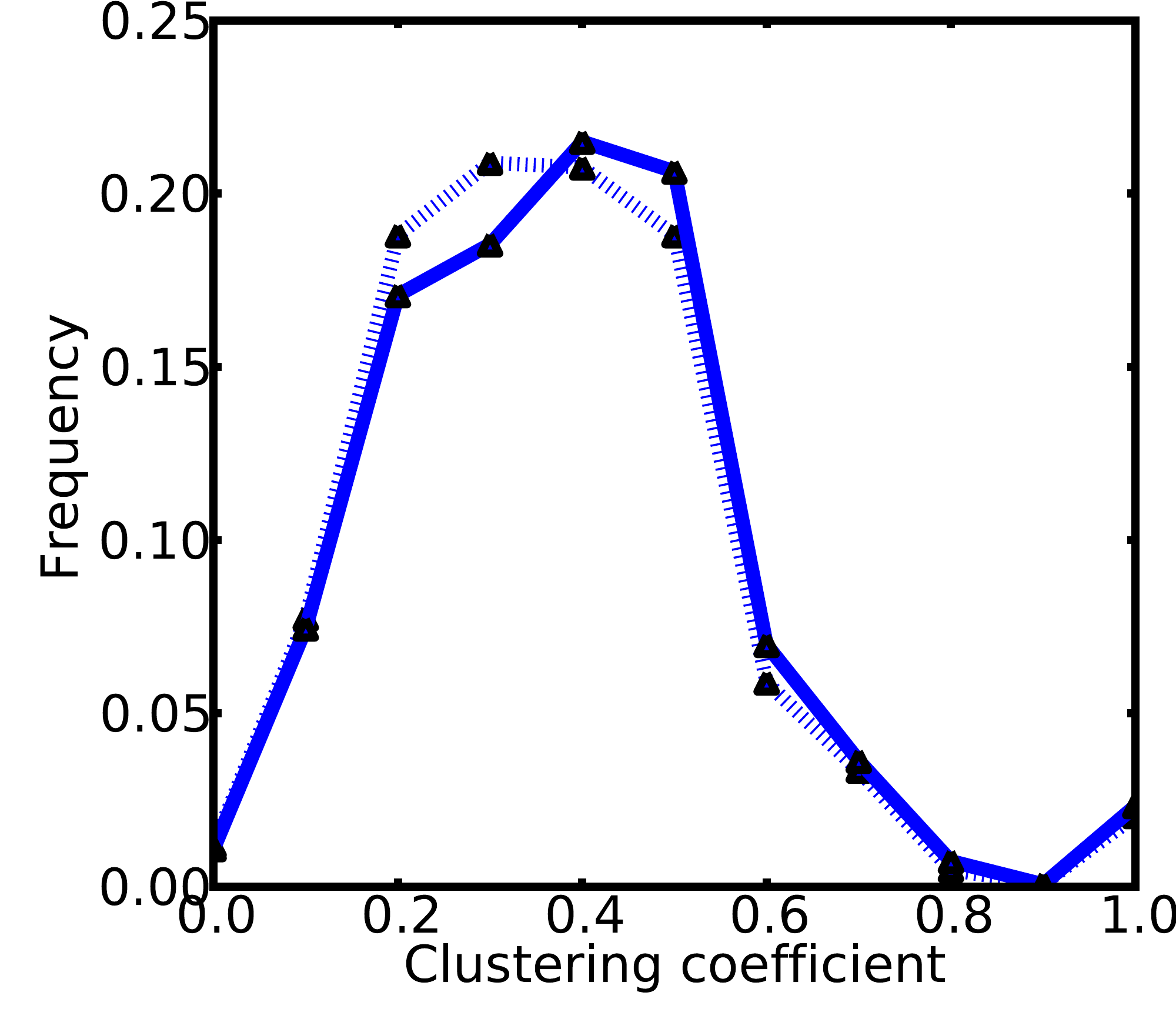} &
    \includegraphics[width=0.5\columnwidth]{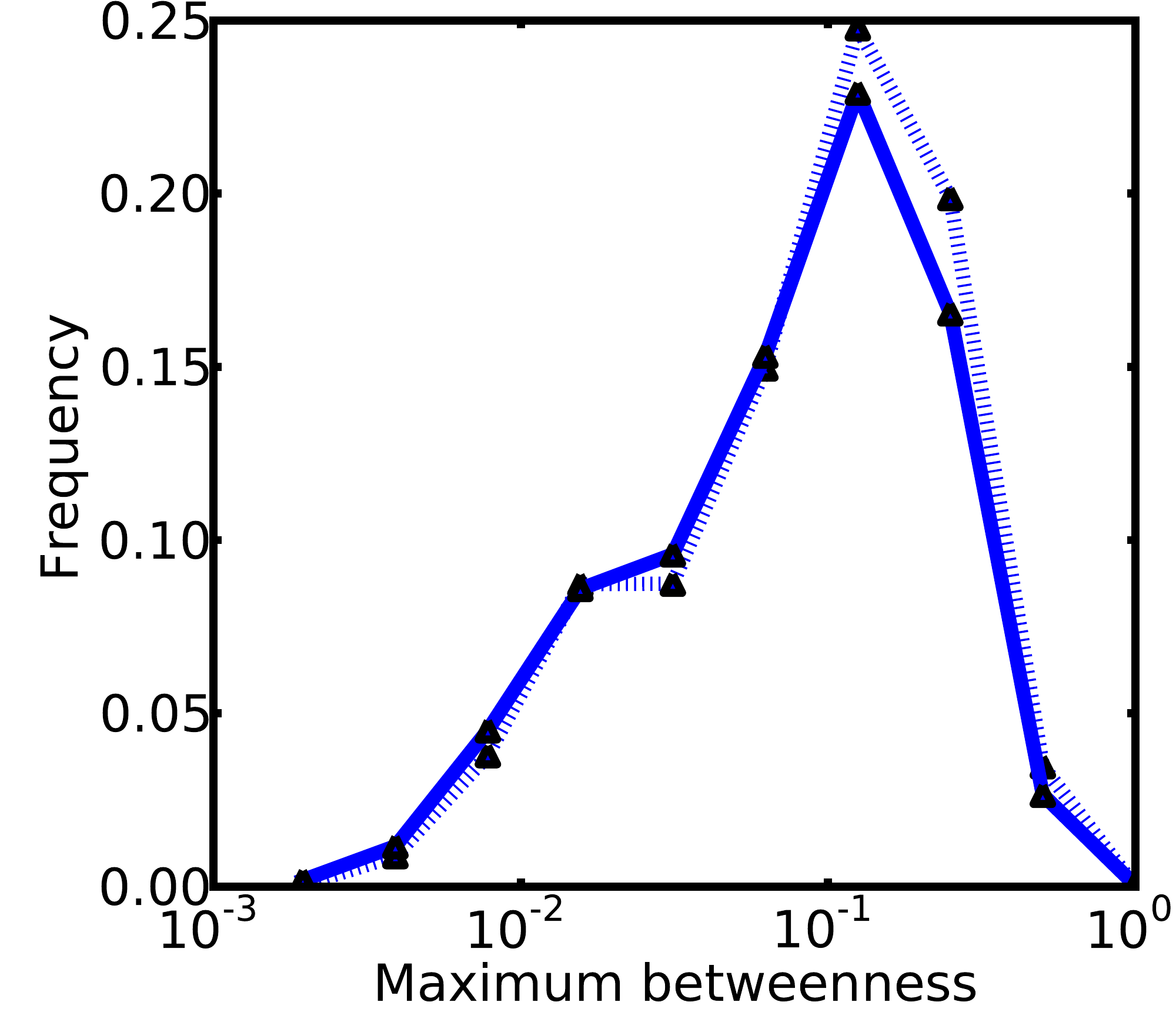} &
    \includegraphics[width=0.5\columnwidth]{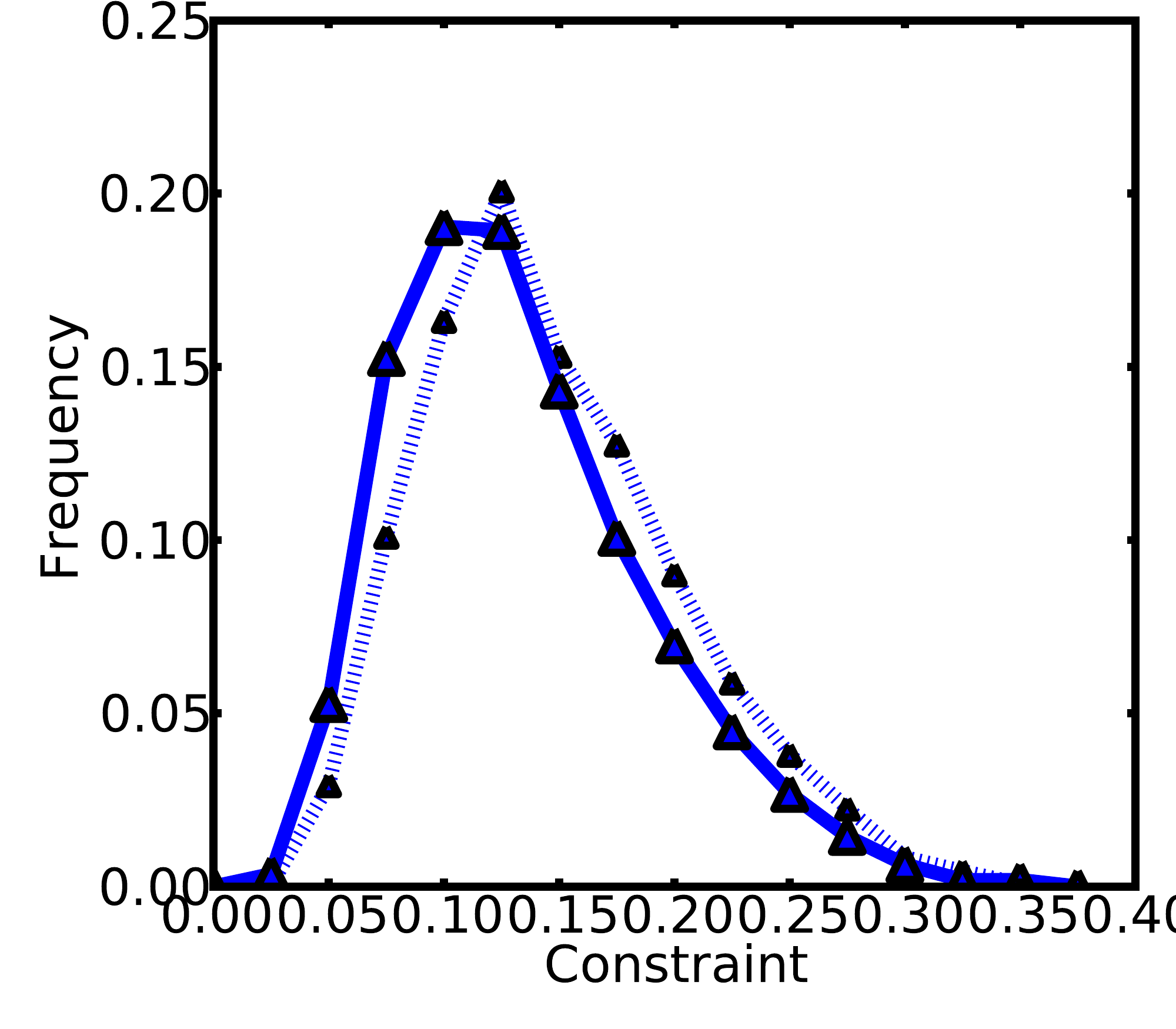}  \\
    (b) M2: Clustering Coefficient & (d) M4: Maximum Betweenness of $v_0$ & (f) M6: Network constraint of $v_0$\\
  \end{tabular*}
  \caption{Citation projection graphs for old and new publications. Notice how science
  is becoming more interdisciplinary as newer papers have lower connectivity and
  significantly smaller network constraint. $t$-tests show that all the differences are statistically significant.}
  \label{fig:overtime}
\vspace{-3mm}
\end{figure*}

First, note that high impact publications have the highest values of connectivity (M3) and maximum betweenness (M4). In all other metrics it is the medium impact publications that have the highest values in the network statistics. This means that in natural and social science it is the medium impact publications that have dense, highly clustered networks with low betweenness and high network constraint of $v_0$. This suggests that medium impact publication in these two areas cite in a very focused and narrow way similar to the illustrative example in Figure~\ref{fig:examples}(b).

Second, note that high and low impact publications have citation networks that are not as dense and clustered as medium impact publications. However, we see that high impact publications have the highest connectivity (M3) and maximum betweenness (M4). In contrast, low impact publications have the lowest connectivity (M3) and the highest betweenness of $v_0$ (M5), which means that shortest paths between the nodes in $G_{p0}$ travel through $v_0$. This suggests that while low and high impact publications tend to both cite a very wide range of sources from different areas and communities, citation graphs of low impact publications are very disconnected (as for example in Figure~\ref{fig:examples}(a)), which indicates idiosyncratic citation behavior. High impact papers also cite a very wide range of papers from a number of communities but they do so in a way that bridges connected areas (high values of M3 and M4). Thus, high impact papers have citation projection graphs that are well connected (high M3) but relatively sparse (medium M1) (similar to the example in Figure~\ref{fig:examples}(c)). This suggests that a high impact paper in natural science or social science is more likely to be an interdisciplinary paper that explores connections between dense but distant fields.

In computer science, the high impact papers have the smallest betweenness of $v_0$ (M5) and highest network density (M1), clustering coefficient (M2), connectivity (M3), maximum betweenness (M3), and network constraint (M6). This means that high impact computer science papers tend to cite publications within single dense communities, which, in turn, suggests that high impact computer science papers tend to be specific and somewhat focused on a single discipline or community inside the larger area of computer science.

All in all we find that high impact publications in natural science and social science tend to be bridges linking cited publications in different clusters of communities together. Although low impact papers also tend to be bridges between areas, citation sub-graphs of high impact papers are better structured (e.g., have higher connectivity and clustering coefficient). This indicates the high risk and high return of citing across communities in natural science and social science~\cite{Shi2009}. However, for the high impact papers the cross community citing is so that there still exists some connection between the communities.

Interestingly, computer science does not seem to follow this pattern. High impact papers in computer science have citation networks that are very focused and do less cross disciplinary citing.

\section{Citation Patterns over Time}\label{sec:time}

In order to gain more insights into the trends in scientific disciplines we analyze how citation practices in various scientific disciplines change over time. We investigate how the properties of citation projection graphs change over time. We divide the publications in natural science into two groups: one includes 15,157 publications from 1900 to 1990, and the other has 14,594 publications from 2000 to 2008. Similarly, for publications in computer science, the group of old publications consists of 6,245 publications from 1920 to 1995, and the group of recent publications is of 5,319 publications from 1996 to 2005.
We now compare the citation projection networks across these two populations of ``old'' and ``recent'' papers.

Figure~\ref{fig:overtime} superimposes the properties of citation projection graphs for ``old'' and ``recent'' papers in natural science and in computer science. We observe very interesting trends. For natural science there is a clear trend of increasing interdisciplinarity of publications, as more recent publications have significantly lower connectivity, lower network constraint and higher betweenness of $v_0$. All these suggest that in natural science recent papers tend to cite a more diverse set of papers. Trends seem to be similar in computer science, however less pronounced. This is due to the fact that the age differences between old and recent papers in the ACM data are small, as most papers in the ACM data appear after 1990.

All these differences between citation projection graphs of old and recent papers suggest that recent scientific publications, especially in the past decade, tend to have broader and more diverse citations than in the past. Moreover, a recent finding shows that electronic access tends to make researchers cite more recent articles in a more focused manner \cite{Evans07182008}. This conclusion is based on the fact that the overall number of articles being cited is getting smaller over time. We note that such analysis ignores the relationships between the cited papers. So, our approach digs deeper into the relationship between the references of individual publications, and shows how the citation relationships in the recent decade differ from those of the past. Together with the finding of \cite{Evans07182008}, it is possible that the electronic access to scientific publications makes researcher focus more on the high profile publications while it also allows them to have a more diverse sample of relevant literature at the same time.

\section{Discussions and Conclusions}\label{sec:conclusion}

In this paper we introduced the notion of citation projection networks and investigated citation behaviors in a large collection of scientific publications spanning natural, social and computer sciences. Our main finding is that there are significant differences in how high, low and medium impact papers position their citations. Whereas medium impact papers tend to create citations in a narrow, well defined and connected field, the citation networks of low and high impact publications are much more diverse and similar to one another. They both cite a very diverse set of sources and refer to publications in various scientific communities. However, the high impact publications are able to find bridges and connections between these scientific communities, whereas the low impact ones are not. Our study indicates the high risk and high return of citing across communities in natural science and social science. In contrast, computer science behaves differently, as high impact papers there tend to have very focused citation networks mostly referring to papers from a narrow community. However, by analyzing temporal trends in citation patterns we found that both natural and computer science are getting more interdisciplinary over time as the citation networks are getting more diverse and increasingly papers from multiple communities are cited.

As a venue for future work, it would be interesting investigate whether citations are more used in network referencing terms, than in labeling terms. By that we mean that scholars may be using citations to reference disciplines as categories (journal names). The heterophily or homophily of citations may also reveal field norms of citation as well as factors contributing to an article's recognition, or citation impact.

\section{Acknowledgements}

We would like to thank JSTOR for providing the article citation data. This project has been generously funded by the Office of the President at
Stanford University and by NSF Award \#0835614. We also would like to thank the generous gifts by Microsoft, Yahoo!, IBM and Lightspeed Ventures.

\bibliographystyle{abbrv}
\bibliography{citations}
\end{document}